**Coping with geometric discontinuities in porous shallow water models**

Giada Varra[1], Renata Della Morte[2], Luigi Cimorelli[3], Luca Cozzolino[4]

**Abstract**

*Porosity-based models are a viable alternative to classical two-dimensional (2-d) Shallow water Equations (SWE) when the interaction of shallow flows with obstacles is modelled. The exact solution of the Single Porosity (SP) Riemann problem, which is the building block of numerous porosity models solved with the Finite Volume method, exhibits an interesting feature, namely the multiplicity of solutions when a supercritical flow impinges on a sudden porosity reduction. In the present paper, this ambiguity is overcome by systematically comparing the solution of the one-dimensional (1-d) SP Riemann problem with the corresponding 2-d SWE numerical solutions at local porosity discontinuities. An additional result of this comparison is that the SP Riemann problem should incorporate an adequate amount of head loss through porosity discontinuities when strongly supercritical flows are considered. An approximate Riemann solver, able to pick the physically congruent solution among the alternatives and equipped with the required head loss amount, shows promising results when implemented in a 1-d Single Porosity Finite Volume scheme.*

[1] Res., Ph. D., Dept. of Engrg., Parthenope Univ., Centro Direzionale di Napoli – Is. C4, 80143 Napoli, Italy. E-mail: giada.varra@uniparthenope.it
[2] Full Prof., Ph. D., Dept. of Engrg., Parthenope Univ., Centro Direzionale di Napoli – Is. C4, 80143 Napoli, Italy. E-mail: renata.dellamorte@uniparthenope.it
[3] Ass Prof., DICEA, Federico II Univ., Via Claudio 21, 80125 Napoli, Italy. E-mail: luigi.cimorelli@unina.it
[4] Ass. Prof., Ph. D., Dept. of Engrg., Parthenope Univ., Centro Direzionale di Napoli – Is. C4, 80143 Napoli, Italy. E-mail: luca.cozzolino@uniparthenope.it

Corresponding author: Giada Varra

E-mail address: giada.varra@uniparthenope.it

Address: Dipartimento di Ingegneria, Università degli Studi di Napoli Parthenope, Isola C4, 80143 Napoli (Italy)

# 1. Introduction

Porosity-based shallow water models have been developed over the last two decades to provide a viable alternative to classical two-dimensional (2-d) Shallow water Equations (SWE) for partially dry areas and transitional environments (Defina 2000), large-scale urban flood modelling (Guinot and Soares-Frazão 2006, Sanders et al. 2008) and runoff simulation on vegetated hillslopes (Ion et al. 2022).

The available porosity shallow water models differ from each other by the conceptual formulation and the underlying physical assumptions. However, Varra et al. (2020) have demonstrated that the Single Porosity (SP) model (Guinot and Soares-Frazão 2006), the Binary Single Porosity model (BSP, Varra et al. 2020), and the integral formulation of the SWE with obstacles by Sanders et al. (2008), constitute a family of models which share the same mathematical structure and features such as hyperbolicity, presence of non-conservative products, and small disturbance celerities coinciding with those exhibited by the 2-d SWE model. Not surprisingly, the observation that a common Finite Volume numerical framework can be used for their approximate solution confirms their common mathematical structure. The Integral Porosity model (IP, Sanders et al. 2008), the Dual Integral Porosity model (DIP, Guinot et al. 2017), and the numerical schemes by Cozzolino et al. (2018b) and Cea and Vázquez-Cendón (2010), are all examples of numerical schemes falling in this framework.

Due to the rapid variations of urban fabric density and flow characteristics through the urban environment, the numerical fluxes over 2-d Finite Volume cell edges are usually calculated by solving a local plane SP Riemann problem (Guinot and Soares-Frazão 2006, Sanders et al. 2008, Finaud-Guyot et al. 2010, Cea and Vázquez-Cendón 2010, Cozzolino et al. 2018b, Jung 2022). The presence of porosity discontinuities requires the definition of appropriate generalized Rankine-Hugoniot conditions (LeFloch 1989, Dal Maso et al. 1995), i.e., relationships between the flow variables at the two sides of the porosity discontinuity that have a strong influence on the Riemann solution.

Numerical methods should incorporate the Rankine-Hugoniot conditions to reproduce the corresponding Riemann exact solutions at porosity discontinuities.

Regarding these solutions, previous studies (Cozzolino et al. 2018a, Varra et al. 2020, 2021) have shown that they present a fundamental ambiguity consisting in the appearance of multiple exact solutions for certain initial conditions characterized by a supercritical flow impacting on a porosity reduction. At this point, two problems arise, namely the need i) to solve this ambiguity by finding the unique physically congruent solution among the alternatives and ii) to construct a numerical scheme able to reproduce the corresponding relevant solution. The one-dimensional (1-d) SP model formally coincides with the 1-d SWE in rectangular channels with variable width, where the porosity symbol substitutes the width symbol (Guinot and Soares-Frazão 2006, Sanders et al. 2008). This physical analogy is exploited in the present paper to disambiguate the multiple 1-d SP Riemann exact solutions by means of a systematic comparison with the corresponding 2-d SWE numerical solutions at local geometric discontinuities, because the 1-d variable-width SWE model is nothing but a crude simplification of the 2-d SWE model in a rectangular channel.

Besides the effects of friction (Guinot et al. 2018), shallow flows in urban environments may dissipate energy by means of different mechanisms, such as the drag induced by obstacles (Sanders et al. 2008), the propagation of bores reflected by buildings (Guinot et al. 2017, 2018), and local effects at geometric discontinuities (Guinot and Soares-Frazão 2006, Varra et al. 2020). In porosity models, the adoption of computational cells of greater size than that usually adopted in shallow water models causes the loss of geometrical and hydraulic information, which in turn causes the underestimation of the energy dissipated by the flow propagating through the urban fabric (Guinot et al. 2017, 2018, Varra et al. 2020). To reproduce missing dissipative effects, structural changes have often been introduced in the original porosity shallow water models, for example altering the physical momentum fluxes via reduction coefficients (Guinot et al. 2017, 2018).

Laboratory (Akers and Bokhove 2008, Defina and Viero 2010) and 2-d SWE numerical experiments (Varra et al. 2020) show that supercritical flows suffer intense head loss across channel contractions, implying that a corresponding energy dissipation must be experienced through rapid porosity reductions. In the present work, this energy dissipation is considered by appropriately reformulating the generalized Rankine-Hugoniot conditions in a head-balance form. The introduction of an interface head loss through the definition itself of porosity discontinuity has the advantage of leaving the structure of the mathematical model unchanged and it can be very naturally used to take into account, at least partly, the drag forces through urban fabrics. However, the amount of head loss to be introduced across the discontinuity needs to be evaluated in a proper way, depending on the flow characteristics across the geometric transition. Also in this case, a systematic study of 2-d SWE numerical results at isolated geometric discontinuities is conducted to supply the general conditions under which this energy dissipation is present and how it can be evaluated.

With the aim of reproducing the effects that in 2-d shallow water models are caused by the flow interaction with isolated geometric discontinuities, the present work proposes a novel approximate Riemann solver that discriminates the existence of multiple solutions and considers adequate head loss in case of supercritical flows at porosity discontinuities. This solver is implemented in a 1-d Finite Volume scheme adopting the Single Porosity formulation of SWE (Guinot and Soares-Frazão 2006). The capability of the 1-d numerical model with porosity of reproducing the effects that in 2-d models are caused by the interaction between the flow and a geometric transition is assessed against several Riemann problems by comparing the corresponding results with the ones provided by a reference 2-d SWE numerical model.

The present paper is organized as follows. The structure of the SP Riemann problem solution, where the definition by Cozzolino et al. (2018b) is used for generalized Rankine-Hugoniot conditions, is discussed in Section 2. This solution is validated in Section 3 using 2-d SWE numerical experiments, and a novel definition of porosity discontinuity is given in Section 4 to better reproduce the 2-d SWE numerical results. In Section 5, it is shown how it is possible to construct a numerical model able to discriminate multiple solutions and introduce the requested head loss amount. These findings are discussed in Section 6. Finally, the paper is closed by a Conclusions section.

## 2. Mathematical model

In the present Section, the plane Riemann problem for the SP model is reviewed, showing that it reduces to a 1-d SP Riemann problem. The corresponding solution requires the definition of generalized Rankine-Hugoniot conditions to be used through porosity discontinuities. Exploiting the analogy between the 1-d SP model and the 1-d SWE in rectangular channels with variable width (Guinot and Soares-Frazão 2006, Sanders et al. 2008), we introduce and discuss a head-balance form defining this relationship.

### 2.1 The 1-d SP model

The plane SP model considered here is an augmented 1-d system obtained from the 2-d SP model (Guinot and Soares-Frazão 2006) by setting to zero the derivatives with respect to the $y$-axis and neglecting the flow resistance components (Ferrari et al. 2017, Varra et al. 2020):

$$(1) \begin{cases} \dfrac{\partial \varphi h}{\partial t} + \dfrac{\partial \varphi h u}{\partial x} = 0 \\ \dfrac{\partial \varphi h u}{\partial t} + \dfrac{\partial}{\partial x}\left( \varphi \dfrac{g h^2}{2} + \varphi h u^2 \right) - \dfrac{g h^2}{2} \dfrac{\partial \varphi}{\partial x} = 0 \\ \dfrac{\partial \varphi h v}{\partial t} + \dfrac{\partial \varphi h u v}{\partial x} = 0 \end{cases}.$$

The solution of the corresponding Riemann problem is the building block for the computation of interface numerical fluxes in shock capturing Finite Volume schemes (Godlewski and Raviart 1996). In Eq. (1), the symbols have the following meaning: $x$ and $y$ are the space independent variables of the inertial reference frame $Oxy$, while $t$ is the time variable; $h(x, y, t)$ is the flow depth; $u(x, y, t)$ and $v(x, y, t)$ are the vertically averaged components of the flow velocity along $x$ and $y$, respectively; $g$ is the gravity acceleration; and the porosity $\varphi(x, y) \in [0, 1]$ represents the fraction of urban area not occupied by buildings and obstacles (storage porosity). In the following, the dependence of the variables on *y* will be omitted because all the quantities should be considered constant along *y* (plane problem). The effects of variable bed elevation are neglected here because the focus of the present work is on obstacle modelling.

Despite a distinction is often made in the urban hydrology literature between storage porosity $\varphi$ and conveyance porosity $\psi$ (Lhomme 2006, Guinot and Delenne 2014, Guinot et al. 2017), the last being related to the mass and momentum transport (Dewals et al. 2021), the two definitions are genuinely different only in porosity models written in integral form while they coincide in differential models (Varra et al. 2020). This result, which derives from a classical proof developed in the theory of fluid motion in porous media (Whitaker 1969), states that the geometric parameter $\varphi$ in Eq. (1) should always be interpreted not only as a storage but also as a conveyance porosity.

In Finite Volume schemes for the approximate solution of the 2-d SP model, the system of Eq. (1), where *x* is a local reference normal to the cell interface, is solved using initial discontinuous conditions (Guinot and Soares-Frazão 2006, Soares-Frazão et al. 2008, Sanders et al. 2008, Cea and Vázquez-Cendón 2010, Finaud-Guyot et al. 2010, Özgen et al. 2016b, Özgen et al. 2017, Guinot et al. 2017). The presence of the non-conservative product $0.5gh^2 \, \partial\varphi/\partial x$, which models the force per unit-width exerted by the obstacles on the flow through the cell interface, requires careful mathematical and numerical treatment because it cannot be recast in divergence form (Cozzolino et al. 2018b). This point is central to the present discussion, and it will be further clarified in the following.

The first two relations of Eq. (1) do not contain the conserved variable *hv* and can be decoupled from the third (Varra et al. 2021), leading to the 1-d SP model (Sanders et al. 2008, Cozzolino et al. 2018b)

(2) $$\frac{\partial \varphi \mathbf{u}}{\partial t} + \frac{\partial \varphi \mathbf{f}(\mathbf{u})}{\partial x} + \mathbf{h}(\mathbf{u})\frac{\partial \varphi}{\partial x} = 0.$$

In Eq. (2), the meaning of the symbols is as follows: $\mathbf{u} = (h \quad hu)^T$ and $\mathbf{f}(\mathbf{u}) = (hu \quad 0.5gh^2 + hu^2)^T$ are the vectors of the conserved variables and fluxes, respectively, of the 1-d SWE model; $T$ is the matrix transpose symbols; $\mathbf{h}(\mathbf{u}) = (0 \quad -0.5gh^2)^T$ is a vector representing the hydrostatic thrust per unit-width exerted by obstacles on the flow.

The 1-d system of Eq. (2) is at the core of the solution to the plane Riemann problem for Eq. (1). In fact, once that $h$ and $hu$ are known from Eq. (2), $hv$ in Eq. (1) is readily computed with the passive tracer equation (Varra et al. 2021):

(3) $\dfrac{\partial v}{\partial t} + u \dfrac{\partial v}{\partial x} = 0$.

### 2.1.1 Porosity Riemann problem

In the Riemann problem of the 1-d SP model, Eq. (2) is solved under the following discontinuous flow initial conditions and porosity

(4) $\mathbf{u}(x,0) = \begin{cases} \mathbf{u}_L, & x < 0 \\ \mathbf{u}_R, & x > 0 \end{cases}$, $\varphi(x) = \begin{cases} \varphi_L, & x < 0 \\ \varphi_R, & x > 0 \end{cases}$,

where $\mathbf{u}_L = (h_L \quad h_L u_L)^T$ and $\mathbf{u}_R = (h_R \quad h_R u_R)^T$ are the states initially to the left and right of the geometric discontinuity in $x = 0$, respectively, while $\varphi_L$ and $\varphi_R$ are the corresponding porosities. The solution of Eq. (2) with the initial conditions of Eq. (4) is self-similar and consists of a sequence of constant states, the leftmost and rightmost of which are $\mathbf{u}_L$ and $\mathbf{u}_R$. All these states are in turn connected by standing or moving waves (Varra et al. 2021). Being the solution self-similar, it exists a vector function $\mathbf{w}(\xi)$ of the scalar parameter $\xi$ such that the Riemann problem solution can be expressed as $\mathbf{u}(x,t) = \mathbf{w}(x/t)$. This implies that the states $\mathbf{u}_1 = (h_1 \quad h_1 u_1)^T$ and $\mathbf{u}_2 = (h_2 \quad h_2 u_2)^T$ immediately to the left and right of the geometric discontinuity, respectively, are constant in time because they can be expressed as $\mathbf{u}_1 = \mathbf{u}(0^-, t) = \mathbf{w}(0^-)$ and $\mathbf{u}_2 = \mathbf{u}(0^+, t) = \mathbf{w}(0^+)$.

Based on the initial conditions of Eq. (4), the porosity is uniform to the left and right of the geometric discontinuity in $x = 0$, implying that the non-conservative product $0.5gh^2 \, \partial\varphi/\partial x$ is null and the system of Eq. (2) is conservative for $x < 0$ and $x > 0$. It follows that the moving waves (shock or rarefactions) coincide with those of the classic 1-d SWE model and that the shocks are defined by the classic Rankine-Hugoniot conditions (Varra et al. 2021). *Vice versa*, the non-conservative product $0.5gh^2 \, \partial\varphi/\partial x$ is active through $x = 0$, implying that the classic Rankine-Hugoniot conditions cannot be used at porosity discontinuities. For this reason, the generalized Rankine-Hugoniot conditions introduced by LeFloch (1989) and Dal Maso et al. (1995) must be used to define an appropriate relationship between $\mathbf{u}_1$ and $\mathbf{u}_2$. In the SP Riemann problem, the self-similarity of the solution implies that this relationship is constant in time because $\mathbf{u}_1$ and $\mathbf{u}_2$ are constant in time.

*2.1.2 Generalized Rankine-Hugoniot conditions at porosity discontinuities*

Following the definition introduced by Dal Maso et al. (1995) for hyperbolic systems of differential equations with non-conservative products, the generalized Rankine-Hugoniot conditions across the porosity discontinuity in Eq. (2) reduce to (Cozzolino et al. 2018b, Varra et al. 2020)

(5.a) $\varphi_R h_2 u_2 - \varphi_L h_1 u_1 = 0$,

(5.b) $\varphi_R \left[ \dfrac{g}{2} h_2^2 + h_2 u_2^2 \right] - \varphi_L \left[ \dfrac{g}{2} h_1^2 + h_1 u_1^2 \right] = S_\Gamma (\varphi_L, \varphi_R, \mathbf{u}_1, \mathbf{u}_2)$,

where $S_\Gamma(\varphi_L, \varphi_R, \mathbf{u}_1, \mathbf{u}_2)$ is the force exerted on the flow by the obstacles across the unit-width porosity discontinuity in $x = 0$. Eq. (5.a) states that the unit-width discharge $Q = \varphi h u$ is invariant through the discontinuity, while Eq. (5.b) states that the total thrusts to the left and right of the discontinuity are balanced by the force $S_\Gamma(\varphi_L, \varphi_R, \mathbf{u}_1, \mathbf{u}_2)$.

The relationship between the states $\mathbf{u}_1$ and $\mathbf{u}_2$ is completely defined if a functional expression for $S_\Gamma(\varphi_L, \varphi_R, \mathbf{u}_1, \mathbf{u}_2)$ is given. However, this expression is somehow problematic because very natural assumptions such as stagnant water and hydrostatic pressure distribution for the computation of $S_\Gamma(\varphi_L, \varphi_R, \mathbf{u}_1, \mathbf{u}_2)$ (Guinot and Soares-Frazão 2006, Sanders et al. 2008, Mohamed 2014, Guinot et al. 2017) may lead to unphysical conditions where the flow acquires energy through the porosity discontinuity (see the discussion in Chow 1959 and Cozzolino et al. 2018b).

To simplify the expression of the relationship between $\mathbf{u}_1$ and $\mathbf{u}_2$, we conveniently reformulate the generalized Rankine-Hugoniot conditions. Appendix A shows that the force balance of Eqs. (5.a) and (5.b) can be rewritten in the following head-balance form

$$\text{(6)} \quad \begin{aligned} & \varphi_R h_2 u_2 - \varphi_L h_1 u_1 = 0 \\ & H(\mathbf{u}_2) - H(\mathbf{u}_1) = \Delta H(\varphi_L, \varphi_R, \mathbf{u}_1, \mathbf{u}_2) \end{aligned},$$

where $H(\mathbf{u}) = h + u^2/(2g)$ is the head corresponding to the generic state $\mathbf{u}$ and $\Delta H(\varphi_L, \varphi_R, \mathbf{u}_1, \mathbf{u}_2)$ is the head loss through the porosity discontinuity. The relationship between the head loss $\Delta H(\varphi_L, \varphi_R, \mathbf{u}_1, \mathbf{u}_2)$ and the force $S_\Gamma(\varphi_L, \varphi_R, \mathbf{u}_1, \mathbf{u}_2)$ is given by (see Appendix A)

$$\text{(7)} \quad \Delta H(\varphi_L, \varphi_R, \mathbf{u}_1, \mathbf{u}_2) = \frac{1}{4}\left[3h_2 + \frac{\varphi_L h_1^2}{\varphi_R h_2} - \frac{\varphi_R h_2^2}{\varphi_L h_1} - 3h_1\right] + \frac{1}{2g}\frac{\varphi_R h_2 + \varphi_L h_1}{\varphi_L \varphi_R h_1 h_2} S_\Gamma(\varphi_L, \varphi_R, \mathbf{u}_1, \mathbf{u}_2),$$

implying that the choice of $S_\Gamma(\varphi_L, \varphi_R, \mathbf{u}_1, \mathbf{u}_2)$ in Eq. (5.b) is equivalent to the choice of $\Delta H(\varphi_L, \varphi_R, \mathbf{u}_1, \mathbf{u}_2)$ in Eq. (6), and *vice versa*.

From the mathematical point of view, the head-balance form is equivalent to the force-balance form, but Eq. (6) is more convenient because it allows to easily verify the physical congruence of $\Delta H(\varphi_L, \varphi_R, \mathbf{u}_1, \mathbf{u}_2)$. In fact, the flow energy cannot increase through the porosity discontinuity, implying that the entropic condition

(8) $Q\Delta H(\varphi_L, \varphi_R, \mathbf{u}_1, \mathbf{u}_2) \leq 0$,

where $Q = \varphi_R h_2 u_2 = \varphi_L h_1 u_1$, must be verified. The head-balance approach allows in principle to easily introduce local effects at geometric discontinuities that do not explicitly appear in Eq. (1), such as non-hydrostatic flow, viscosities, velocity variability along the vertical direction, flow depth and velocity variability along the transverse directions, and the shape of obstacles. Not surprisingly, existing definitions of channel internal boundary conditions such as width discontinuities and junctions from the technical literature are usually given in terms of head loss (Formica 1955, Austin et al. 1970, Cunge et al. 1980, Hager 2010).

However, the system of Eq. (2) does not provide additional information to compute the head loss $\Delta H(\varphi_L, \varphi_R, \mathbf{u}_1, \mathbf{u}_2)$, implying that the hydraulic modeller should use external physical knowledge for its definition. This will be discussed in the following subsection, where the internal structure of the porosity discontinuity between $x = 0^-$ and $x = 0^+$ will be examined.

## 2.2 Channel analogy and porosity discontinuity definition

The 1-d SP model of Eq. (2) coincides with the 1-d SWE model in a rectangular channel with variable width and horizontal bed, where the porosity symbol $\varphi$ takes the place of the width symbol $B$ (Guinot and Soares-Frazão 2006, Sanders et al. 2008, Varra et al. 2021). Like the porosity models written in differential form, where the storage and conveyance porosity must coincide, the width $B$ in rectangular channels can be regarded both as i) the channel base-area per unit length (storage) and ii) the transverse space available for the flow (conveyance). In addition, the non-conservative product $0.5gh^2 \partial\varphi/\partial x$ in Eq. (2), which represents the force per unit-width exerted by the obstacles on the flow along $x$, acts like the corresponding term in the 1-d variable-width SWE model, where it represents the force exerted on the flow by the channel walls. In the following, this channel analogy will be exploited to supply a convenient expression for the head loss $\Delta H(\varphi_L, \varphi_R, \mathbf{u}_1, \mathbf{u}_2)$ through the porosity discontinuity.

### *2.2.1 Porosity variation through the discontinuity*

Consider the bottom of Figure 1a, where a strip of unitary width modelling a simplified urban area with obstacles is depicted. The strip is subdivided into two cells, left and right, respectively, with different obstacle densities represented by the porosities $\varphi_L$ and $\varphi_R$. Obstacles are also present through the cell interface in $x = 0$, with a density intermediate between those of the two adjacent cells. The corresponding rectangular channel analogue is represented by the top channel of Figure 1a, where the widths at the left and right ends are represented by $\varphi_L$ and $\varphi_R$, while a monotonic width variation (channel contraction or expansion) connects the left and right reaches. In this case, it is natural to assume that the porosity $\varphi$ through the discontinuity between $x = 0^-$ and $x = 0^+$ is described by a monotonic function $\varphi_\Gamma(s)$ in the interval $s \in [0, 1]$, with $\varphi_\Gamma(0) = \varphi_L$ and $\varphi_\Gamma(1) = \varphi_R$. In Figure 2a, the internal structure of the porosity discontinuity between $x = 0^-$ and $x = 0^+$ is exploded to show the relationship between $\varphi_\Gamma(s)$ and the parameter $s \in [0,1]$.

A similar situation is depicted in Figure 1b, but now the obstacles density at the cells interface is greater than those of the two adjacent cells. The corresponding rectangular channel analogue is represented by the top channel of Figure 1b, where a non-monotonic width variation (channel constriction) connects the left and right reaches and the porosity $\varphi_\Gamma(s)$ through the discontinuity varies non-monotonically between $\varphi_L$ and $\varphi_R$.

**Remark 1**. In Finite Volume schemes, a homogeneous porosity is assigned to each computational cell and the actual obstacles distribution along the interface between two contiguous cells is canceled. This implies that the simplest application of the channel analogy is to consider a rectangular channel that is normal to the cell interface and symmetrical with respect to its longitudinal axis (as made in Figures 1 and 2).

Thanks to the channel analogy, both the monotonic and non-monotonic choices of $\varphi_\Gamma(s)$ are physically viable and lead to numerically stable computations. Examples of models with monotonic porosity variation through the discontinuity are contained in the works by Guinot and Soares-Frazão (2006), Soares-Frazão et al. (2008), Cea and Vázquez-Cendón (2010), Finaud-Guyot et al. (2010), Ferrari et al. (2017), and Cozzolino et al. (2018a,b), while examples with a non-monotonic variation are the models by Sanders et al. (2008), Özgen et al. (2017), Bruwier et al. (2017), and Guinot et al. (2017, 2018, 2022).

The preceding discussion suggests that the choice of the porosity discontinuity internal structure can be made considering the underlying urban geometry at each cell interface. This simple observation, which supplies a physically congruent framework, contradicts the unproven statement from the literature (Bruwier et al. 2017, Guinot et al. 2017, 2018, 2022) that only non-monotonic descriptions of the porosity variation through the discontinuity are viable and stable (see the corresponding discussion in Varra et al. 2020).

**Remark 2**. For the sake of simplicity, only monotonic porosity variations through the discontinuity with $\varphi_L \leq \varphi_R$ will be considered in the rest of the paper (monotonic variations with $\varphi_L > \varphi_R$ can be discussed by simply mirroring the local reference framework). Having defined the aspect ratio $AR = \varphi_L/\varphi_R$, this implies that $AR \leq 1$ will be assumed in the following developments.

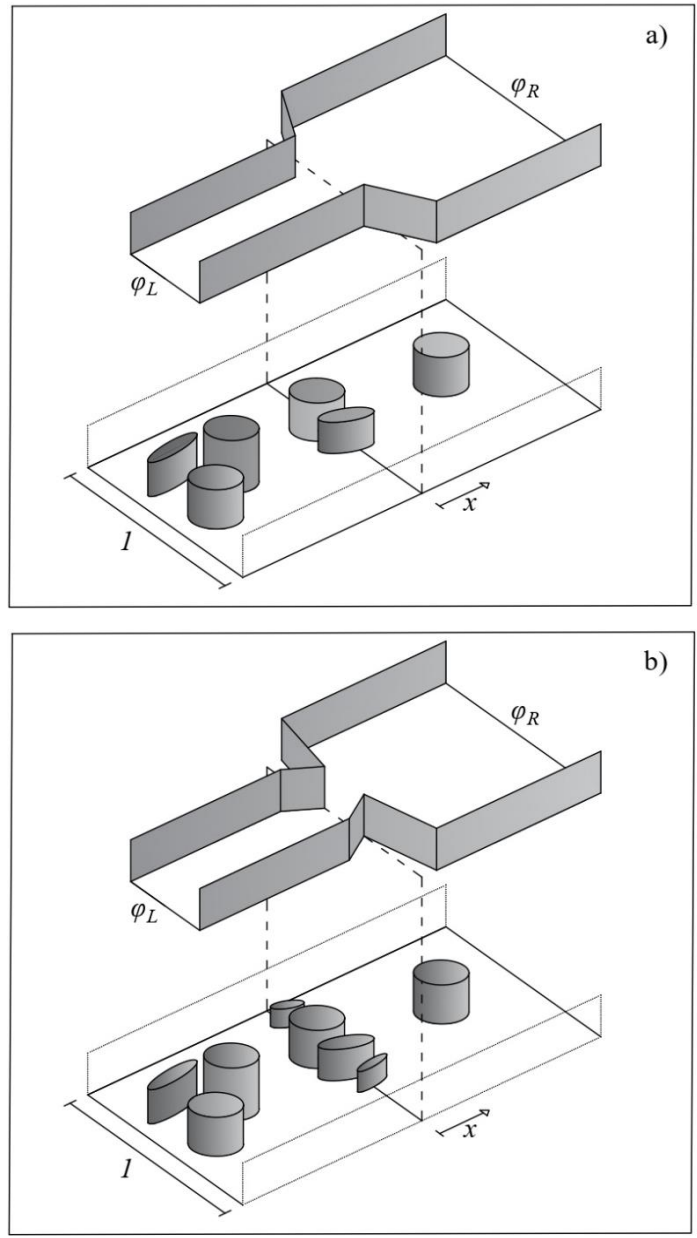

**Figure 1**. Physical interpretation of the porosity discontinuity between $\varphi_L$ and $\varphi_R$: monotonic (a) and non-monotonic porosity variation (b).

### *2.2.2 Flow depth and velocity variation through the discontinuity*

To complete the internal description of the porosity discontinuity, it is necessary to specify the variation of flow depth and velocity between $x = 0^-$ and $x = 0^+$. It is assumed that this description is supplied by the function $\mathbf{v}(s) = (h_\Gamma \quad h_\Gamma u_\Gamma)^T$, where $h_\Gamma(s)$ and $u_\Gamma(s)$, with $s \in [0, 1]$, are the flow depth and velocity through the discontinuity, respectively. The function $\mathbf{v}(s)$ is characterized by the obvious congruency conditions $\mathbf{v}(0) = \mathbf{u}_1$ and $\mathbf{v}(1) = \mathbf{u}_2$. In Figures 2b and 2c, the internal structure of the porosity discontinuity between $x = 0^-$ and $x = 0^+$ is exploded to show two examples of the relationship between the inner flow depth $h_\Gamma(s)$ and the parameter $s \in [0,1]$.

Cozzolino et al. (2017) have proposed that the function $\mathbf{v}(s)$ is a stationary weak solution of Eq. (2) through the porosity discontinuity, namely a solution of

$$(9) \quad \frac{d\varphi_\Gamma \mathbf{f}(\mathbf{v})}{ds} + \mathbf{h}(\mathbf{v})\frac{d\varphi_\Gamma}{ds} = 0$$

in the interval $s \in [0,1]$. If the solution of Eq. (9) exhibits no hydraulic jump (see the example of Figure 2b, where $h_\Gamma(s)$ smoothly varies in the interval $s \in [0,1]$), the relationship between the states $\mathbf{u}_1$ and $\mathbf{u}_2$ reduces to the conditions of discharge and total head invariance (see Appendix B)

$$(10) \quad \begin{array}{l} \varphi_R h_2 u_2 - \varphi_L h_1 u_1 = 0 \\ H(\mathbf{u}_2) - H(\mathbf{u}_1) = 0 \end{array},$$

which is equivalent to set $\Delta H(\varphi_L, \varphi_R, \mathbf{u}_1, \mathbf{u}_2) = 0$ in the head-balance form of Eq. (6). In the case that the porosity varies monotonically between $\varphi_L$ and $\varphi_R$, the existence of a state $\mathbf{u}_1$ connected to $\mathbf{u}_2$ by means of Eq. (10) depends on the aspect ratio $AR \leq 1$ and on $|F(\mathbf{u}_2)|$, where $F(\mathbf{u}) = u/\sqrt{gh}$ is the Froude number related to the generic state $\mathbf{u}$. The corresponding discussion is reported in Appendix C.

If the solution of Eq. (9) exhibits a hydraulic jump that reverts the incoming supercritical flow into subcritical (see the example of Figure 2c), the total head is not invariant through the porosity discontinuity and the corresponding head loss depends on the position of the hydraulic jump (see Appendices C and D in Varra et al. 2021). The corresponding relationship between the states $\mathbf{u}_1$ and $\mathbf{u}_2$, which recovers the head-balance form of Eq. (6) with $Q\Delta H(\varphi_L, \varphi_R, \mathbf{u}_1, \mathbf{u}_2) < 0$, is not as simple as that of Eq. (10) and it is not reported here for the sake of brevity.

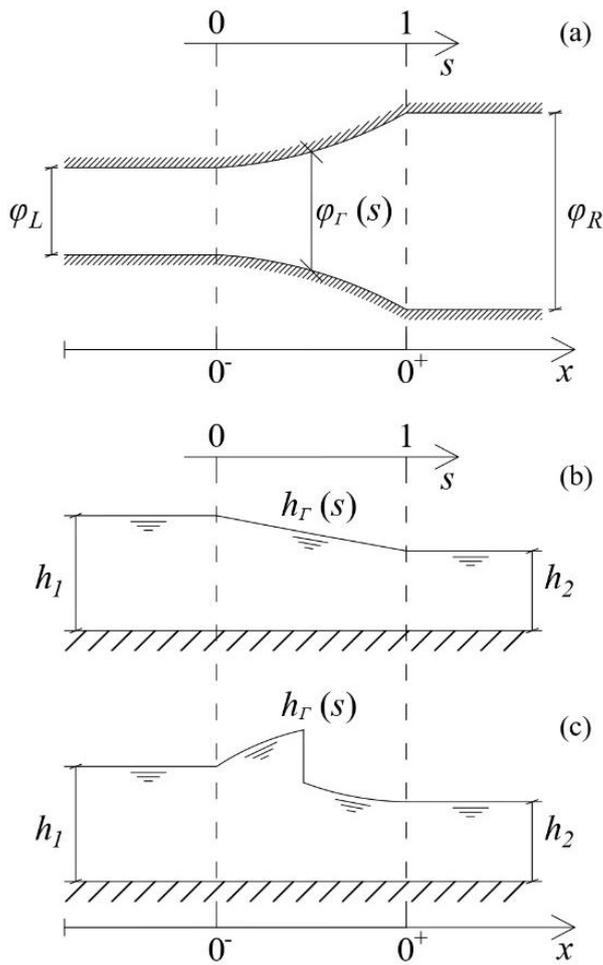

**Figure 2**. Internal description of the porosity discontinuity: plan view of the monotonic porosity variation (a); profile view of smooth flow depth variation (b); profile view of flow depth variation with hydraulic jump (c).

### 2.2.3 Definition of the Generalized Rankine-Hugoniot conditions

Having discussed the internal structure of the porosity discontinuity in the preceding sections, we assume the following definition for the generalized Rankine-Hugoniot conditions:

**Definition 1**. The relationship between $\mathbf{u}_1$ and $\mathbf{u}_2$, with $AR = \varphi_L/\varphi_R \leq 1$, is defined by the head-balance form of Eq. (6) with the following internal description of the porosity discontinuity:

D1) the porosity varies monotonically between $\varphi_L$ and $\varphi_R$;

D2) the variation of flow depth and velocity through the porosity discontinuity is defined by a weak solution of Eq. (9) in the interval $s \in [0,1]$, with $\mathbf{v}(0) = \mathbf{u}_1$ and $\mathbf{v}(1) = \mathbf{u}_2$.

This definition automatically satisfies the entropic condition of Eq. (8), while this is not true for other porosity discontinuity definitions (Guinot and Soares-Frazão 2006, Sanders et al. 2008, Mohamed 2014, Guinot et al. 2017) available in the literature (see the discussion in Cozzolino et al. 2018b). In addition, Varra et al. (2021) have demonstrated that the solution to the Riemann problem of Eqs. (2) and (4) always exists if Definition 1 is used to establish the generalized Rankine-Hugoniot conditions. This fundamental result is not granted for alternative porosity discontinuity definitions from the literature.

**2.3 Multiple solutions to the porosity Riemann problem**

The solution to the 1-d SP Riemann problem of Eqs. (2) and (4), complemented by the generalized Rankine-Hugoniot conditions of Section 2.2.3, always exists but there are cases, depending on the initial conditions $\mathbf{u}_L$ and $\mathbf{u}_R$, where the solution is triple (Varra et al. 2021). The field of occurrence of multiple solutions will be explored in the following for the case $AR = \varphi_L/\varphi_R \leq 1$ only (a similar discussion for the case $AR > 1$ can be drawn by mirroring the reference framework).

The necessary (but not sufficient) condition for the existence of multiple solutions to the 1-d SP Riemann problem with $AR \leq 1$ is that the right state $\mathbf{u}_R$ is directed from right to left ($u_R < 0$) and $|F_R| > K_{sp}(AR)$, where $F_R = F(\mathbf{u}_R)$ is the Froude number corresponding to $\mathbf{u}_R$ while the function $K_{sp}(AR)$ is defined in Appendix C. In this condition, the right supercritical flow $\mathbf{u}_R$ impinging the porosity reduction has energy greater than the minimum required to pass through the geometric discontinuity. For this reason, it is possible to consider not only a solution where the right flow freely passes through the discontinuity, but also solutions where the head of the incoming flow is partially dissipated by means of a standing hydraulic jump through the porosity transition or by a shock that moves backwards (Viero and Defina 2017, Varra et al. 2021)

The theoretical limit curve $|F_R| = K_{sp}(AR)$, called lower boundary (LB) of the hysteresis domains (Viero and Defina 2017), is represented in the plane ($|F_R|$, $AR$) of Figure 3 with a black continuous line. The necessary condition $|F_R| > K_{sp}(AR)$ for the existence of multiple solutions is satisfied by the points falling in the regions denoted with B and C to the right of the LB curve in Figure 3. The regions B and C are separated by the curve $|F_R| = K_{jump}(AR)$, called upper boundary (UB) of the hysteresis domains (Viero and Defina 2017), which is represented with a dashed line. The dimensionless function $K_{jump}(AR)$, which is characterised by $K_{jump}(AR) \geq K_{sp}(AR)$ for every $AR \leq 1$, is defined in Appendix C.

In the regions B and C, one or three different solutions to the 1-d SP Riemann problem of Eqs. (2) and (4) are possible, depending on the initial left state $\mathbf{u}_L$ (Varra et al. 2021). When $\mathbf{u}_L$ is such that three alternative solutions (here called T1, T2, and T3) are possible, the solutions differ from each other by the flow condition through the porosity discontinuity, as follows:

(T1) the state $\mathbf{u}_2$ immediately to the right of the porosity discontinuity coincides with the supercritical flow $\mathbf{u}_R$, while the state $\mathbf{u}_1$ is supercritical and it is connected to $\mathbf{u}_2$ by means of Eq. (10), namely by the conditions of discharge and total head invariance across the discontinuity (Figure 4a);

(T2) the state **u**₂ coincides with **u**_R but a hydraulic jump is present through the porosity discontinuity, and the state **u**₁ is subcritical or critical, with $H(\mathbf{u}_1) < H(\mathbf{u}_2)$ (Figure 4b);

(T3) the supercritical flow **u**_R is reverted into the subcritical state **u**₂ by means of a backward moving shock, with head loss; the state **u**₁ is subcritical or critical and it is connected to **u**₂ by means of Eq. (10) (Figure 4c).

When **u**₁ is subcritical in the solutions T2 and T3, the flow through the geometric discontinuity is submerged, i.e., it is dominated by the tailwater $h_1$. The main difference between the regions B and C in Figure 3 is the behaviour of the solutions to the Riemann problem when the state **u**_L coincides with the dry bed, i.e., when $h_L = 0$ and there is no tailwater. In the domain B, three distinct solutions (T1, T2, and T3) are always possible for $h_L = 0$, while a unique solution T1 occurs in the domain C for $h_L = 0$. In other words, a triple solution is possible in the domain B even if there is not a downstream tailwater able to force the establishing of a subcritical flow through the geometric discontinuity (submerged flow), while such a tailwater is required for the existence of a triple solution in the field C. In a sense, the incoming flow falling in region C has always energy sufficient to flush the hydraulic jump of Figure 4b out of the porosity discontinuity when the flow depth downstream is null.

The discussion is completed observing that the region A to the left of the curve LB, characterised by $1 \leq |F_R| \leq K_{sp}(AR)$, refers to flow conditions where the Riemann problem always admits a unique solution. In this case, the supercritical flow **u**_R has not sufficient energy to pass through the porosity reduction and the solution T3 must occur.

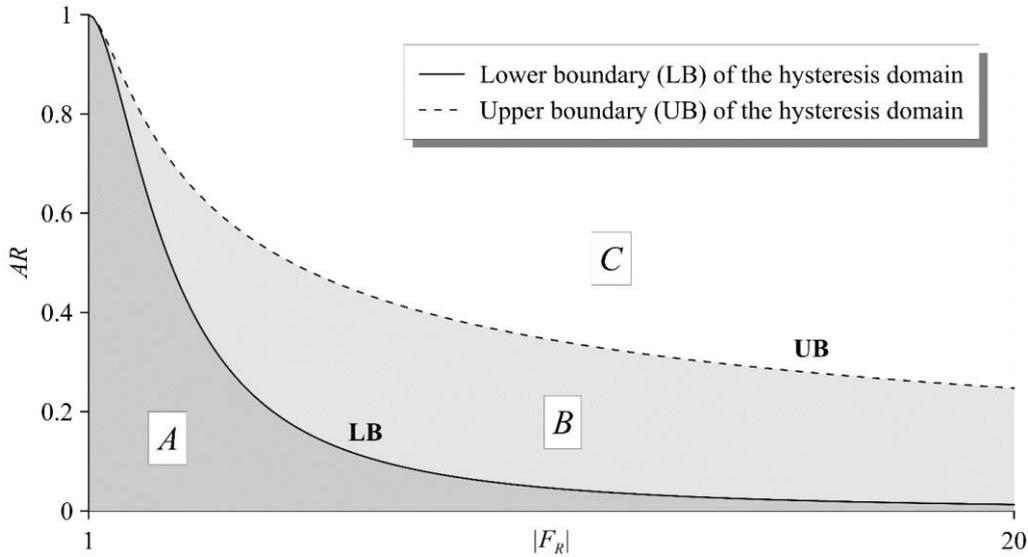

**Figure 3**. Field of occurrence of multiple solutions to the porosity Riemann problem for right supercritical flows $\mathbf{u}_R$ impinging a porosity reduction with $AR = \varphi_L/\varphi_R \leq 1$. Lower (continuous line) and upper (dashed line) boundaries of the hysteresis domains. Hysteresis domains: A (no multiple solutions), B (multiple solutions even in the case $h_L = 0$), C (multiple solutions only for $h_L > 0$).

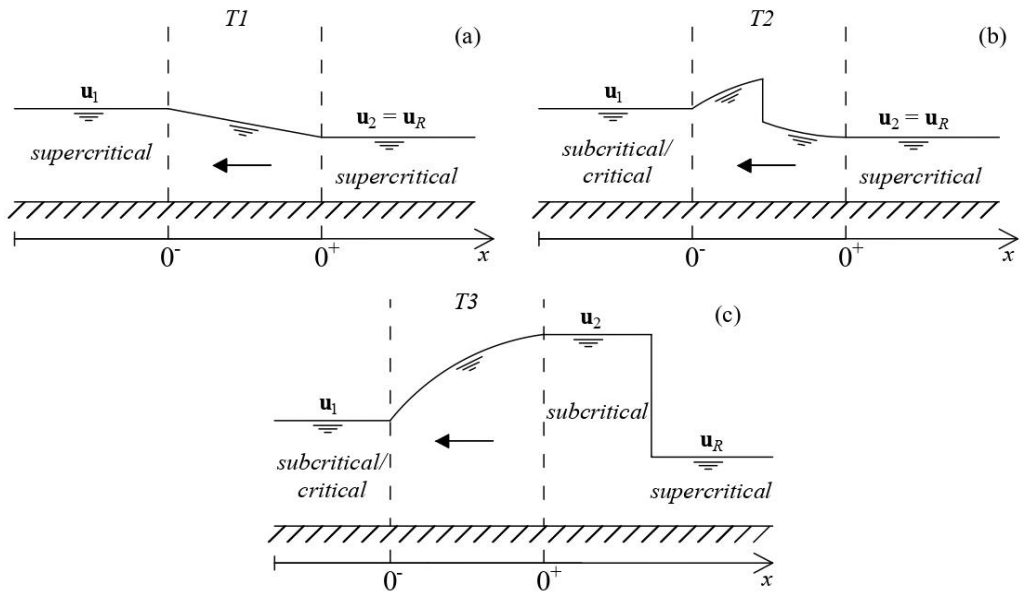

**Figure 4**. Flow conditions through the porosity discontinuity when multiple solutions to the purely 1-d SP Riemann problem are possible: profile view of solutions T1 (a), T2 (b) and T3 (c).

## 3. Validation of the channel analogy

The comparison between 1-d SP and 2-d SWE solutions is justified because the 1-d SP Riemann problem is the main ingredient of 2-d SP Finite Volume schemes, which in turn are intended to approximate the solution of 2-d SWE models with obstacles. For this reason, the generalized Rankine-Hugoniot conditions of Section 2.2.3 are validated in the present section by comparing several 1-d SP exact Riemann solutions with the corresponding 2-d SWE numerical solutions in a frictionless horizontal rectangular channel with variable width, where a 2-d contraction is used to model the 1-d sudden porosity reduction. All the Riemann problems, whose initial conditions $\mathbf{u}_L$ and $\mathbf{u}_R$ with the corresponding Froude numbers $F_L$ and $F_R$ are reported in Table 1, refer to porosity values $\varphi_L = 0.6$ and $\varphi_R = 1$. The exact Riemann solutions are computed with the methods discussed in Varra et al. (2021).

**Table 1**. Initial flow conditions of the validation Riemann problems.

| Example | $h_L$ (m) | $u_L$ (m/s) | $F_L$ (-) | $h_R$ (m) | $u_R$ (m/s) | $F_R$ (-) |
|---|---|---|---|---|---|---|
| 1 | 1.00 | 2.00 | 0.64 | 1.00 | -0.50 | -0.16 |
| 2 | 1.00 | 2.00 | 0.64 | 1.00 | 2.00 | 0.64 |
| 3 | 1.00 | 5.00 | 1.60 | 1.00 | 2.00 | 0.64 |
| 4 | 0.30 | -10.00 | -5.83 | 1.00 | 2.00 | 0.64 |
| 5 | 1.00 | -2.00 | -0.64 | 1.00 | -9.40 | -3.00 |
| 6 | 1.00 | 7.00 | 2.23 | 1.00 | -13.00 | -4.15 |
| 7 | 1.00 | -11.00 | -3.51 | 1.00 | -13.00 | -4.15 |
| 8 | 0.30 | -4.00 | -2.33 | 0.30 | -11.00 | -6.41 |

The rectangular channel considered for the 2-d SWE computations has length $L = 200$ m with a left reach of width $B_L = 0.60$ m and a right reach of width $B_R = 1.00$ m (see Figure 5). The left and right channel reaches are separated by a symmetric linear expansion whose length is $L_c = 0.20$ m. The 2-d SWE computations are accomplished using the Finite Volume scheme described in Cozzolino et al. (2017) on unstructured triangular grid whose average side is $\Delta s = 0.50$ m at the channel ends and $\Delta s = 0.05$ m at the linear expansion. The flow depth $h$ computed at time $t = 5$ s with the 1-d SP exact solution (continuous black line) is compared in Figures 6, 8, 9, and 10, with the corresponding 2-d shallow water numerical results (white dots).

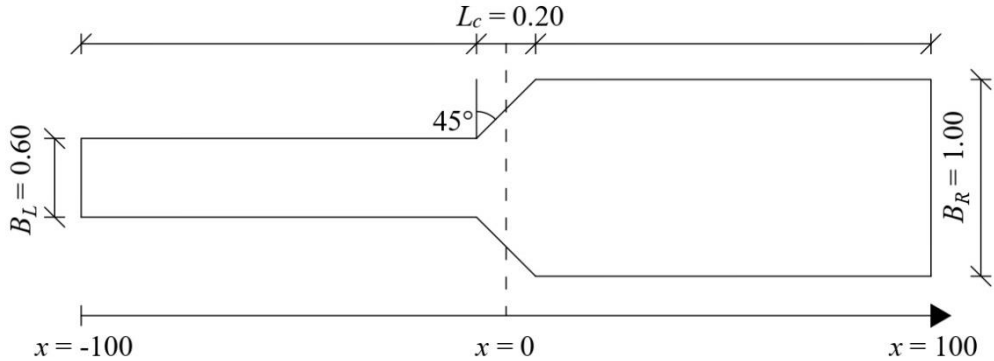

**Figure 5**. Plan view of the channel considered for 2-d SWE numerical simulations. Distorted representation (measures in metres).

The results of Riemann problem 1 are represented in Figure 6a. The 1-d exact solution presents two shocks moving to the left and right of the geometric discontinuity, respectively, while subcritical flow conditions that preserves discharge and energy invariance are established through $x = 0$. Figure 6a shows a good correspondence between the 1-d exact and 2-d numerical solutions. In particular, the 1-d model accurately captures the strength of the wave at $x = 0$, together with the strength and position of the shocks. The intermediate states $\mathbf{u}_1$ and $\mathbf{u}_2$ immediately to the left and right of the porosity discontinuity computed with the 1-d exact solution nicely correspond to those provided by the 2-d SWE model.

A slightly different picture can be drawn for the solution of Riemann problem 2, represented in Figure 6b. The 1-d exact solution exhibits a resonant condition where a rarefaction is attached to the left of the porosity discontinuity, while a shock and a rarefaction are both moving to the right. The state $\mathbf{u}_1$ immediately to the left of $x = 0$ is critical and accelerates through the rapid geometric transition becoming supercritical with preservation of energy and discharge invariance. In turn, the supercritical state $\mathbf{u}_2$ issuing from the channel expansion pushes the slowly moving shock to the right of $x = 0$. With reference to the left rarefaction, Figure 6b shows a good correspondence between the 1-d exact and 2-d numerical results. This representation is less satisfactory with reference to the right moving shock, whose shape in the 2-d model is strongly influenced by its vicinity to the channel expansion. Similarly, the 2-d rarefaction moving on the right is quite smoothed with respect to the 1-d exact solution. Despite these discrepancies, the intermediate state between the shock and the rarefaction computed with the 1-d exact solution satisfactorily corresponds to the 2-d numerical solution.

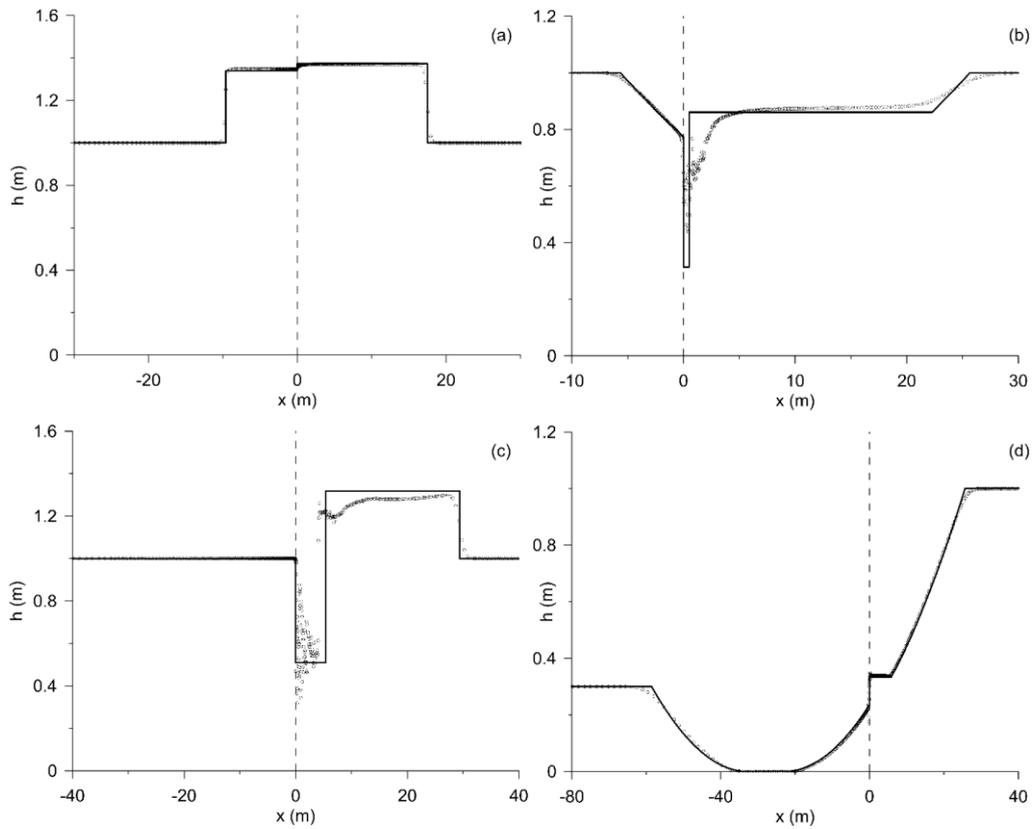

**Figure 6**. Profile view of the 1-d SP exact (continuous black line) and 2-d SWE numerical solutions (dots) for the flow depth at time $t = 5$ s. Example Riemann problems 1 (a), 2 (b), 3 (c) and 4 (d) with initial conditions in Table 1.

In the 1-d exact solution of Riemann problem 3, the supercritical states $\mathbf{u}_1$ and $\mathbf{u}_2$ are connected by the conditions of discharge and head invariance while the state $\mathbf{u}_2$ is separated from $\mathbf{u}_R$ by means of an intermediate state and two moving shocks. The comparison with the 2-d SWE results shows that the strength and position of the right moving shock and the left shock position are satisfactorily captured by the 1-d exact solution, while the flow depth of the state between the two shocks supplied by the1-d model is not too far from the 2-d solution. The 2-d free surface profile view in Figure 6c exhibits a complex pattern whose plane view is represented in Figure 7, which shows that the supercritical flow accelerating through the expansion originates a system of transverse oblique shocks. The 1-d exact solution represents this pattern with a single average flow depth, and this explains the discrepancies between 1-d and 2-d models. Nonetheless, the 1-d model captures the general picture of the 2-d SWE solution.

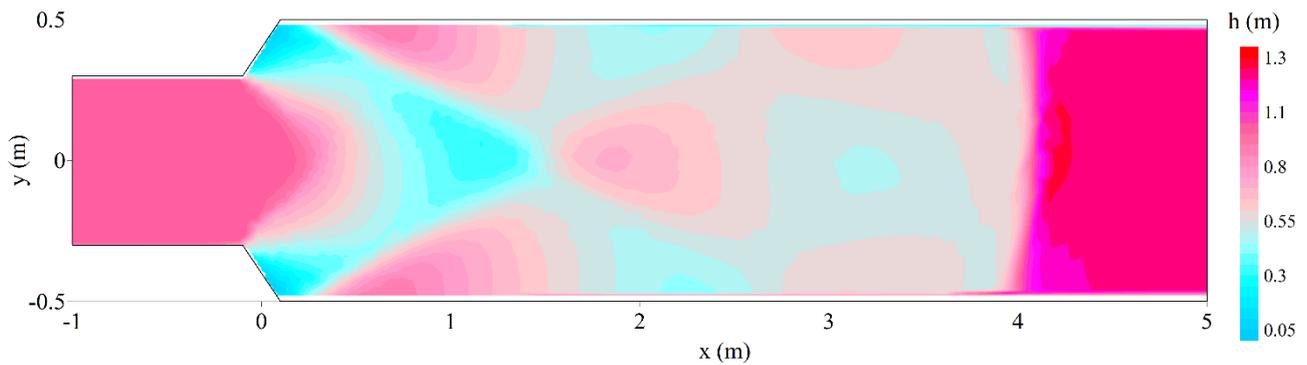

**Figure 7**. Plan view of the 2-d SWE numerical solution for Riemann problem 3 with initial conditions in Table 1. Flow depth contours at time $t = 5$ s.

In Figure 6d, the results of Riemann problem 4 are represented. The 1-d exact solution consists of two rarefactions to the left of $x = 0$, with the formation of dry bed between the waves, and of an additional rarefaction to the right of the discontinuity. The subcritical flow immediately to the right of $x = 0$ (state $\mathbf{u}_2$) accelerates through the discontinuity becoming critical immediately to the left (state $\mathbf{u}_1$) and preserving the invariance of discharge and energy. The comparison between the 1-d exact and 2-d SWE numerical results shows that the former captures the strength of the 2-d waves, together with the flow depth of the states encompassed by the waves.

The example of Figure 8 is particularly interesting because it corresponds to Riemann problem 5 with three exact solutions, where $AR$ and $|F_R|$ fall in the hysteresis domain B of Figure 3. In Figure 8a, b, c, the three exact solutions T1, T2, and T3, respectively, are represented (see Section 2.3). In Figure 8d, the superposition between the exact solution T3 and the 2-d SWE numerical results shows a good agreement. This demonstrates that the under-determination of the 1-d Riemann problem can be eliminated by resorting to the 2-d SWE model, which takes into account the transverse flow variability.

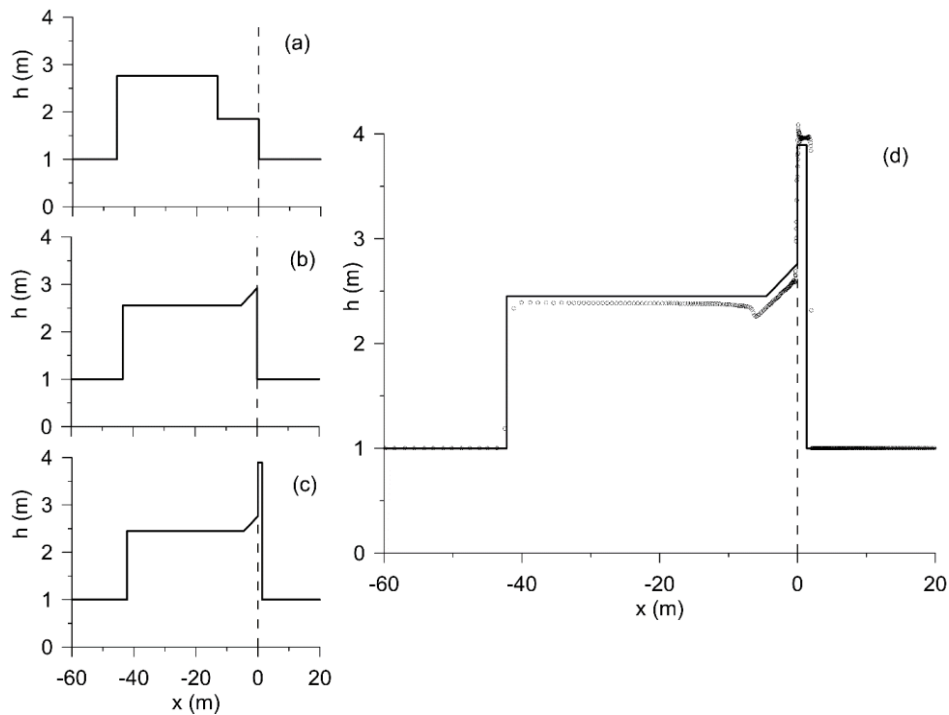

**Figure 8**. Example Riemann problem 5 with initial conditions in Table 1. Profile view for the flow depth solution at time $t = 5$ s. 1-d SP exact solutions T1 (a), T2 (b) and T3 (c). Comparison between the T3 exact solution (continuous line) and the 2-d SWE numerical solution (dots) (d).

The example of Figure 9 corresponds to Riemann problem 6 with three exact solutions, where $AR$ and $|F_R|$ fall into the hysteresis domain C of Figure 3. In Figure 9a, b, c, the three exact solutions T1, T2, and T3, respectively, are represented (see Section 2.3). In Figure 9d, the superposition between the T3 exact solution and the 2-d SWE numerical results shows again a good agreement. The comparison between figures 8d and 9d shows that the 2-d SWE model preferably picks up the solution with a shock moving backwards when three exact solutions are possible.

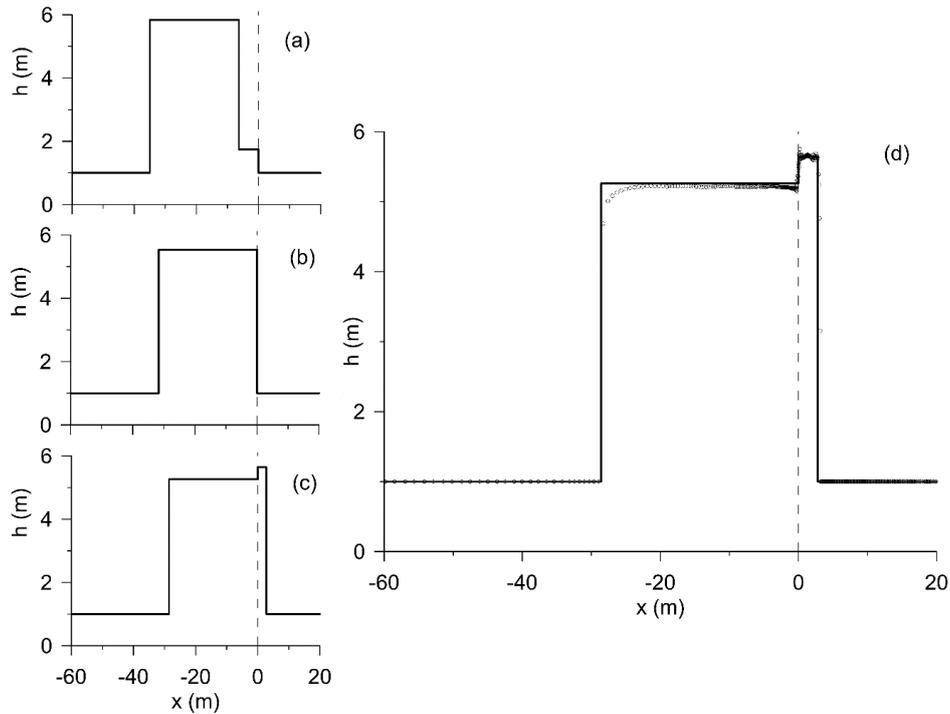

**Figure 9.** Example Riemann problem 6 with initial conditions in Table 1. Profile view for the flow depth solution at time $t = 5$ s. 1-d SP exact solutions T1 (a), T2 (b) and T3 (c). Comparison between the T3 exact solution (continuous line) and the 2-d SWE numerical solution (dots) (d).

In the example of Figure 10a, the state $\mathbf{u}_L$ is such that Riemann problem 7 has one exact solution of type T1, despite $AR$ and $|F_R|$ fall in the hysteresis domain C of Figure 3. Correspondingly, the 2-d SWE solution is characterised by a strong interaction with the geometric discontinuity and by a supercritical flow to the left of $x = 0$. The comparison between 1-d and 2-d solutions shows that the number of moving waves supplied by the 1-d exact solution is correct, but their strength and position is very different from those of the 2-d numerical solution. In addition, the flow depth of the supercritical states to the left of $x = 0$ is poorly captured by the 1-d exact solution. The 2-d SWE numerical solution exhibits a strong interaction with the channel walls in $x = 0$, with the formation of a system of oblique shocks whose plan view is represented in Figure 11. These shocks, which are typical of supercritical flows in contractions (Ippen and Dawson 1951, Akers and Bokhove 2008, Defina and Viero 2010), introduce intense head loss that explains the discrepancies between 1-d and 2-d solutions.

Similar observations can be made for Riemann problem 8, for which $AR$ and $|F_R|$ fall in the hysteresis domain C of Figure 3. The 1-d exact solution is unique and characterized by flow conditions T1 through the porosity reduction (Figure 10b), while the corresponding 2-d SWE solution exhibits a supercritical flow to the left of $x = 0$ and a strong interaction with the contraction (Figure 12) where intense head loss is produced.

From the validation process above, some considerations can be made. The exact solutions to the 1-d SP Riemann problem, computed with the monotonic porosity discontinuity model of Section 2.2.3, compare well with the corresponding 2-d SWE numerical solutions in case of subcritical flow through the porosity discontinuity (Figures 6a,d). Overall, the head loss through the discontinuity seems negligible when the flow is subcritical, but a *caveat* to this observation will be discussed in the next Section.

Similarly, the 1-d exact model shows a good behaviour in both the multiplicity domains B and C when three exact solutions are possible. In this case, the solution characterized by subcritical flow through the porosity discontinuity satisfactorily agrees with the 2-d SWE numerical results (Figures 8d and 9d).

A minor discrepancy is present when the supercritical flow accelerates through a porosity increase. In this case, the 2-d SWE numerical solution is somehow distorted with respect to the 1-d exact solution (Figures 6b,c).

A very different picture is evident when the 1-d exact model predicts a single solution characterized by supercritical flow through a porosity reduction (Figures 10a,b). In this case, the 2-d SWE model exhibits a supercritical flow through the contraction, but the head loss introduced by a

2-d system of oblique shocks makes the 1-d and 2-d solutions very different. This dissipative mechanism has been discussed by Varra et al. (2020) for the first time in the context of Riemann problems on dry bed, but it has been verified here for the general Riemann problem on wet bed.

The validation process accomplished in this section suggests that the results of a 2-d SWE numerical model could be systematically used to disambiguate multiple Riemann problem solutions and evaluate the head loss caused by a supercritical flow passing through a porosity reduction, improving the Definition 1 of the generalized Rankine-Hugoniot conditions. This will be made in the next section.

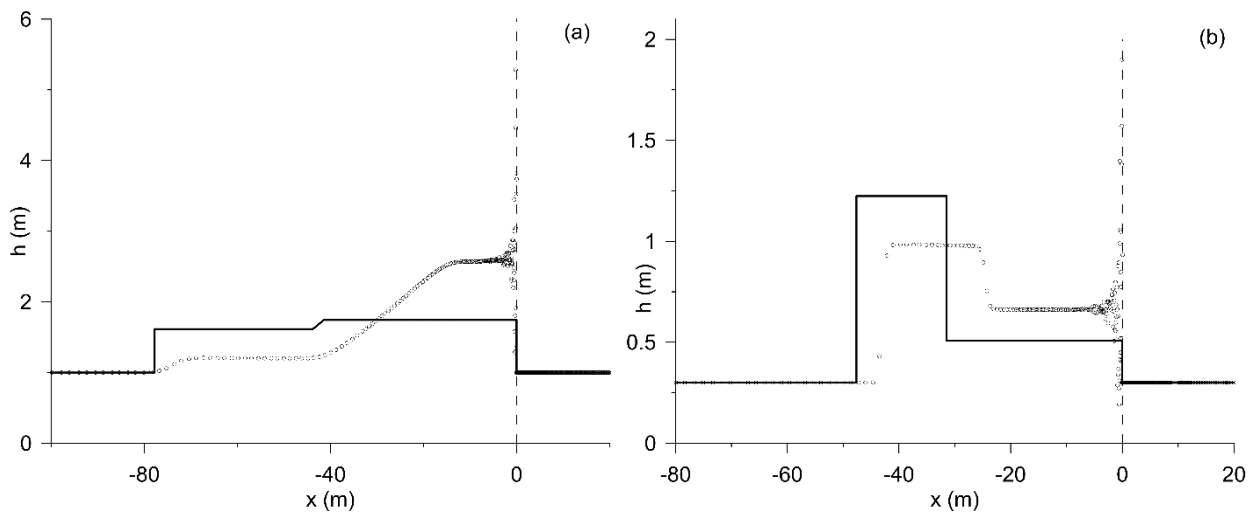

**Figure 10**. Profile view of the 1-d SP exact (continuous black line) and 2-d SWE numerical solutions (dots) for the flow depth at time $t = 5$ s. Example Riemann problems 7 (a) and 8 (b) with initial conditions in Table 1.

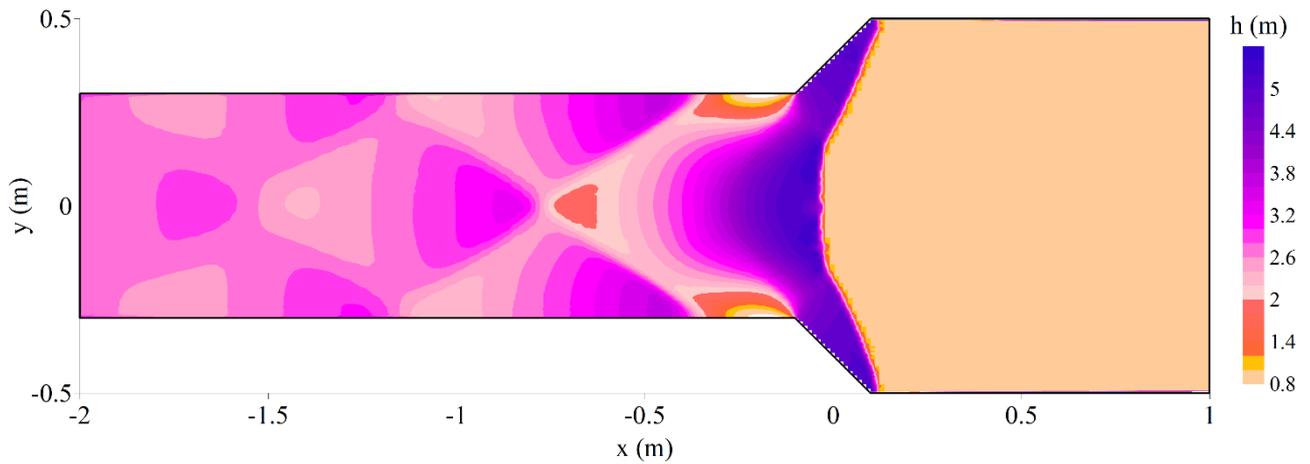

**Figure 11**. Plan view of the 2-d SWE numerical solution for Riemann problem 7 with initial conditions in Table 1. Flow depth contours at time $t = 5$ s.

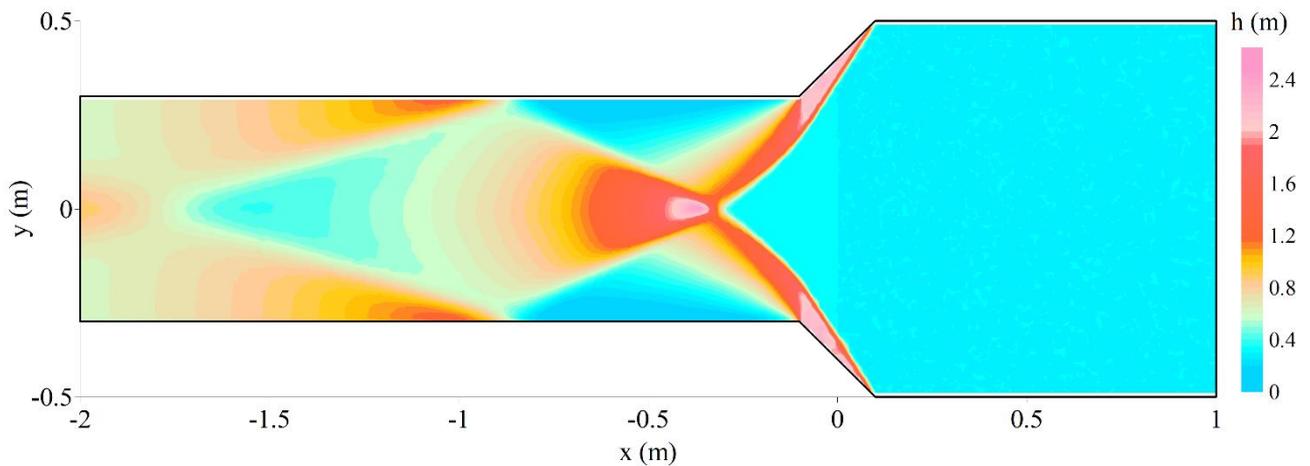

**Figure 12**. Plan view of the 2-d SWE numerical solution for Riemann problem 8 with initial conditions in Table 1. Flow depth contours at time $t = 5$ s.

## 4. Construction of novel generalized Rankine-Hugoniot conditions

Consider a horizontal frictionless rectangular channel $L = 60$ m long with a single width discontinuity at its centre. The channel consists of a right and a left reach of different widths, connected by a linear contraction whose walls are inclined by 45° with respect to the channel axis (see Figure 13). The contraction is short enough to be regarded as a true geometric discontinuity. In order to perform 2-d SWE simulations with different aspect ratio $AR = B_L/B_R$ values, the right reach width is fixed to $B_R = 1$ m, while the left reach width $B_L$ (with $B_L < B_R$) is varied in each test. Free-slip boundary conditions are imposed at the channel walls, while the left and right ends are open. A non-uniform unstructured triangular mesh is used for simulations, with average side $\Delta s = 0.20$ m at the channel ends and $\Delta s = 0.02$ m in the vicinity of the geometric transition.

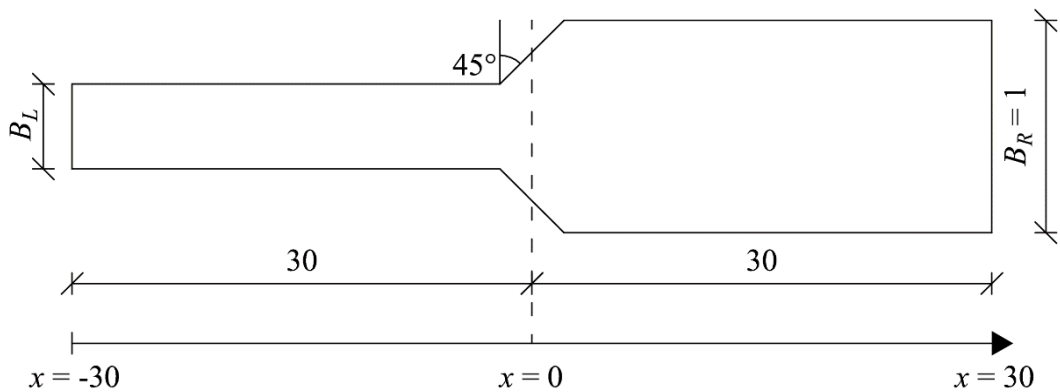

**Figure 13**. Plane view of the channel used for 2-d SWE numerical tests with supercritical flows. Distorted representation (measures in metres).

The 2-d Finite Volume SWE numerical model by Cozzolino et al. (2017) is used for approximating the solution of 159 different 2-d Riemann problems in the channel of Figure 13, where $AR \in [0.1, 0.9]$. The tests are characterized by supercritical flow $\mathbf{u}_R$ (with $u_R < 0$) approaching the contraction and Froude number $|F_R| \in [1.5, 25]$. In all the tests, the right flow depth is $h_R = 1$ m and the corresponding velocity $u_R$ varies accordingly to $F_R$, while the initial left state $\mathbf{u}_L$ coincides with the dry bed. Simulations are run until steady state conditions are reached through the contraction (generally after $t = 20$ s).

The 1-d SP exact solution is triple (T1, T2, or T3) for the initial conditions falling in region B of Figure 3, while a single T1 solution is predicted for points falling in region C. The examination of numerical results shows that two distinct types of 2-d SWE solutions occur:

(G1) The supercritical flow entering the geometric discontinuity passes with the formation of a complicate system of oblique shocks through the contraction like in Figures 10a,b; this set of numerical solutions exhibits a behavior that is somehow in between the exact solutions T1 and T2 defined in Section 2.3 (Figures 4a,b), because supercritical flow conditions are present at the outlet of the contraction as in T1, but there is also head loss as in T2.

(G2) The flow through the contraction is subcritical, while a moving shock propagates upstream; since the channel bed downstream is initially dry, critical flow conditions are established through the contraction outlet; this set of numerical solutions clearly recalls the exact solutions of type T3 in Section 2.3 (Figure 4c), where $\mathbf{u}_1$ is critical.

**4.1 Modified upper boundary of the hysteresis domain**

In Figure 14, the 2-d numerical cases corresponding to solution types G1 (black triangles) and G2 (white squares) are plotted in the plane ($|F_R|$, $AR$), where the upper hysteresis domain limit is also represented. Figure 14 shows that, for a given value of the aspect ratio $AR$, it exists a limit Froude number $K^*(AR)$ such that a G2 solution is obtained for $|F_R| \leq K^*(AR)$, while a G1 solution is obtained for $|F_R| > K^*(AR)$. The locus of the points separating the fields of G1 and G2 solutions is the modified upper boundary curve with equation $|F_R| = K^*(AR)$. Ideally, this curve represents the situations for which a standing hydraulic jump is present at the entrance of the contraction. For $|F_R| > K^*(AR)$, the incoming flow has energy sufficient to push the jump through the contraction, where it is broken into a complicate pattern of transverse standing waves (Figures 11 and 12). Conversely, the incoming flow is not able to sustain the hydraulic jump for $|F_R| < K^*(AR)$, and a shock moves backwards.

The modified upper boundary curve, represented in Figure 14 with a thick black line, is very close to the UB curve defined in Section 2.3 (dashed line curve in Figure 14) for moderate width jumps ($AR > 0.5$), whereas it departs from the UB curve for strong width jumps ($AR \leq 0.5$). This is not surprising, because the 1-d theory for width and porosity transitions is expected to work better when $AR$ is close to one. The polynomial interpolation of data supplies for the limit $|F_R| = K^*(AR)$ the expression

$$(11) \quad K^*(AR) = K_{jump}(AR) \sum_{i=1}^{6} m_i AR^i ,$$

whose coefficients $m_i$ are reported in Table 2.

We observe that $K^*(AR) > K_{jump}(AR)$ for $AR > 0.5$, meaning that the incoming flow requires greater energy to push the hydraulic jump through the porosity discontinuity with respect to the case without energy loss. This effect, which is due to the modest head loss related to the subcritical flow through the geometric transition in G2 solutions, will be taken into account numerically without a direct evaluation of the energy losses in subcritical conditions.

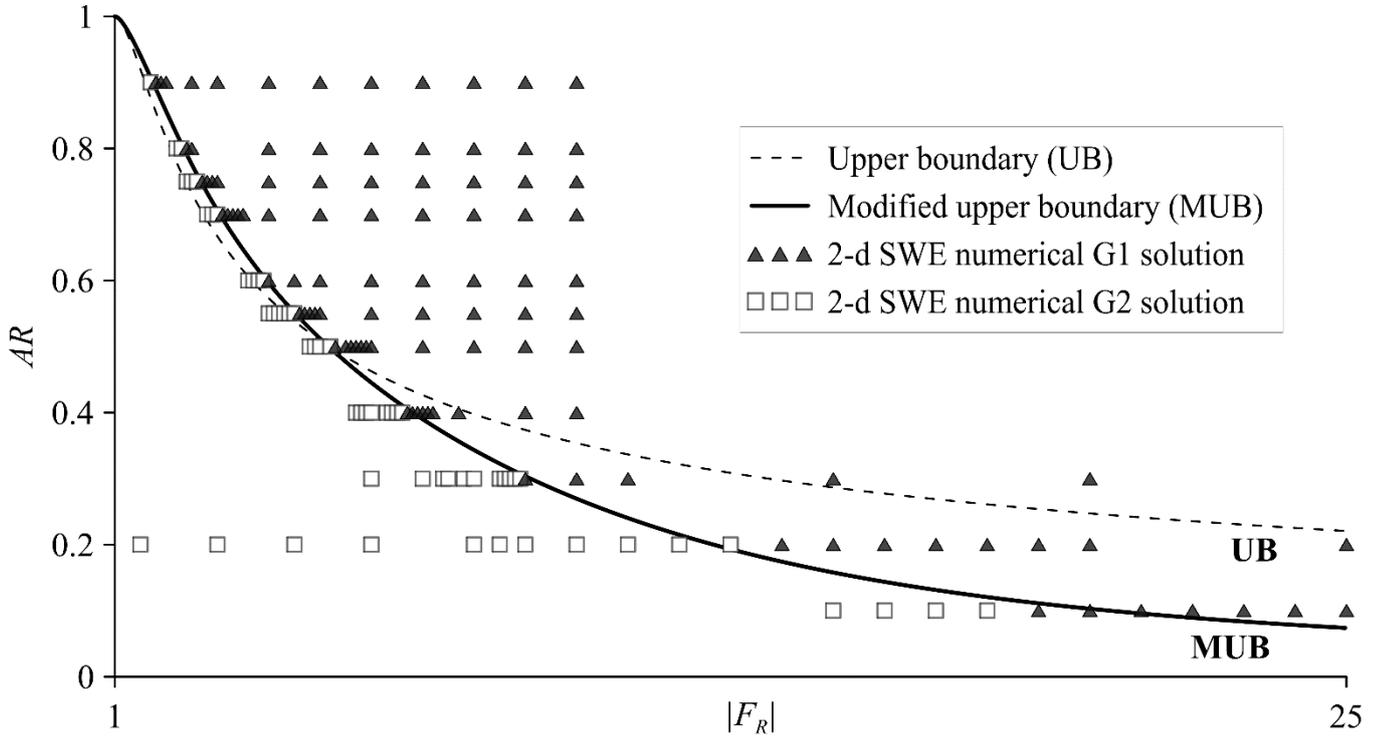

**Figure 14**. 2-d SWE numerical results for supercritical flows with $B_R = 1$ m and $h_R = 1$ m impinging a contraction: G1 configuration (black triangles), G2 configuration (white squares); upper hysteresis domain limit (dashed line); modified upper boundary (thick black line).

**Table 2**. Coefficients for the polynomial interpolation of Eq. (11).

| $m_1$ | $m_2$ | $m_3$ | $m_4$ | $m_5$ | $m_6$ |
|---|---|---|---|---|---|
| 0.9448 | 9.8030 | -24.2944 | 20.1172 | -3.7583 | -1.8122 |

**4.2 Head loss for supercritical flows at contractions**

The 2-d SWE solutions of type G1 exhibit a head loss $\Delta H^*$ through the channel contraction. This head loss is evaluated as $\Delta H^* = H(\mathbf{u}_R) - H_1^*$, where $H_1^*$ is an estimate of the head corresponding to the state immediately to the left of the contraction. The quantity $H_1^*$ is indirectly deduced by evaluating the supercritical flow depth to the left of the contraction.

The relative head loss $\Delta^* = \Delta H^* / H(\mathbf{u}_R)$ corresponding to the 2-d SWE solutions of type G1 is represented with black triangles in the plane ($\Delta^*, F_R^2$) of Figure 15, where the data corresponding to the same *AR* value are connected by a dashed line. Figure 15 shows that the relative head loss $\Delta^*$ moderately varies with $F_R^2$ for a given value of *AR*. This implies that $\Delta^*$ mainly depends on the characteristics of the geometric transition, allowing to use a simplified expression in the form $\Delta^* = \Delta^*(AR)$, where a single $\Delta^*$ value has been attributed to each *AR* value by picking the numerical data closer to the modified upper boundary of Figure 14. These points are connected by a continuous grey line in Figure 15.

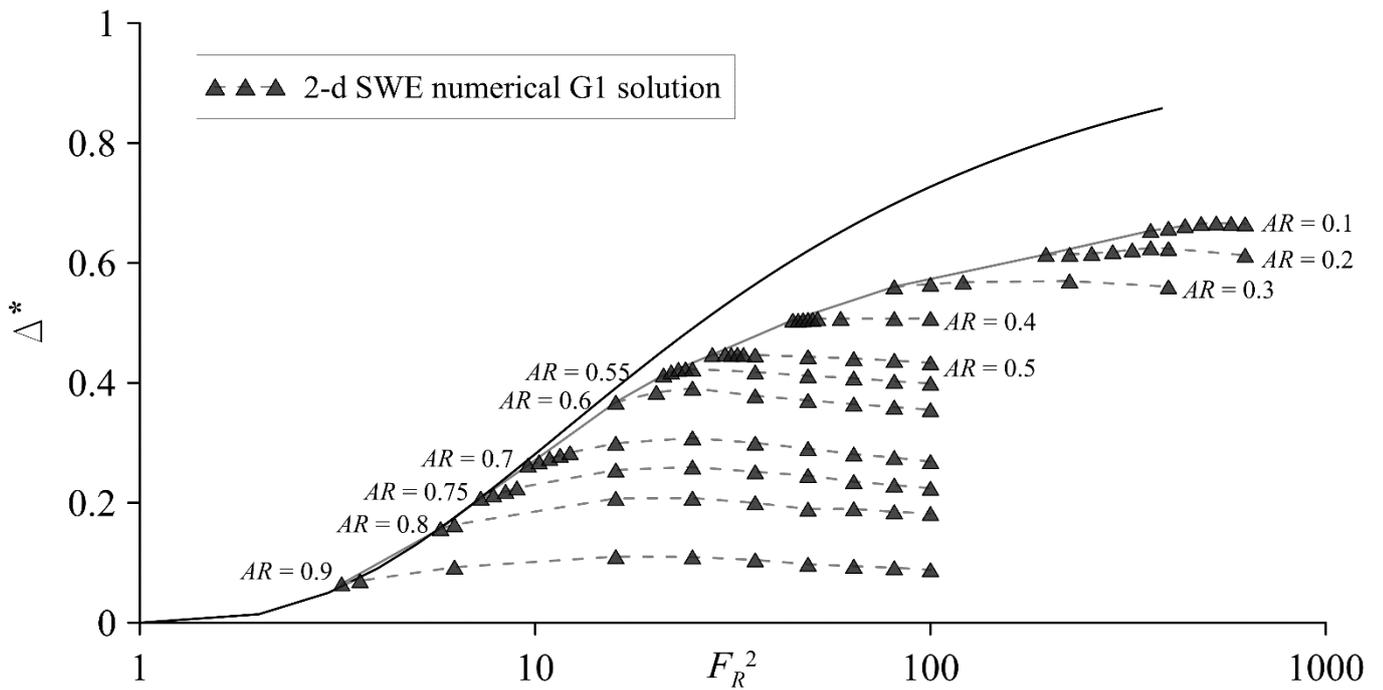

**Figure 15**. Relative head losses for supercritical flows through a contraction: 2-d SWE numerical results for G1 configuration (black triangles); limit relative head loss (continuous black line); envelope of G1 data closer to the modified upper boundary of Figure 14 (continuous grey line).

In the same figure, the limit relative head loss $\Delta^{\#} = \Delta H^{\#}/H(\mathbf{u}_R)$ is also represented with a thick black line. The quantity $\Delta H^{\#}$ is the head loss through the shock in the exact solution of type T3 (see Figure 4c) when the celerity of the shock is null (standing hydraulic jump) and the state $\mathbf{u}_1$ is critical, i.e., when $|F_R| = K_{jump}(AR)$. In Appendix D, it is shown that the limit relative head loss $\Delta^{\#}$ depends on $AR$ only and the exact expression of $\Delta^{\#} = \Delta^{\#}(AR)$ is given. The ratio $\Delta^{*}/\Delta^{\#}$ is represented in Figure 16 for different values of $AR^2$. This leads to the following polynomial interpolation

(12) $\Delta^{*}(AR) = \Delta^{\#}(AR) \sum_{i=0}^{2} m_i AR^{2i}$,

with interpolation coefficients in Table 3.

Recalling that $\mathbf{u}_2 = \mathbf{u}_R$ in the G1 solutions, from the position $\Delta H(\varphi_L, \varphi_R, \mathbf{u}_1, \mathbf{u}_2) = \Delta H^{*}$ it follows that the head loss in Eq. (6) for supercritical flow through a channel contraction, equivalent to a porosity reduction, can be rewritten as

(13) $\Delta H(\varphi_L, \varphi_R, \mathbf{u}_1, \mathbf{u}_2) = H(\mathbf{u}_2) \Delta^{*}(\varphi_L/\varphi_R)$.

Table 3. Coefficients for the polynomial interpolation of Eq. (12).

| $m_0$ | $m_1$ | $m_2$ |
|---|---|---|
| 1.536 | 0.403 | 0.668 |

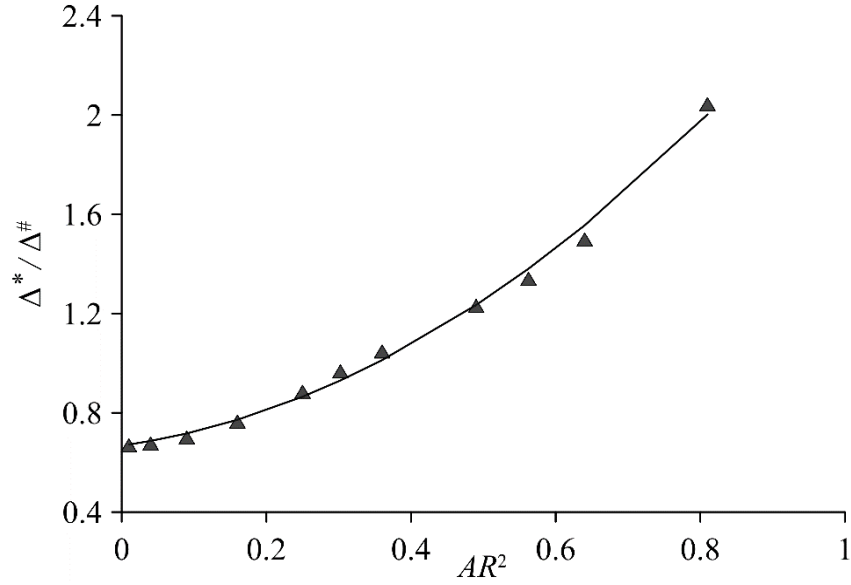

**Figure 16**. Polynomial interpolation of the relative head loss data. Triangles represent the experimental cases enveloped by a thin grey line in Figure 15.

**4.3 Novel generalized Rankine-Hugoniot conditions**

From the preceding discussion, it is possible to give a novel convenient relationship between $\mathbf{u}_1$ and $\mathbf{u}_2$ at porosity discontinuities.

**Definition 2**. The relationship between $\mathbf{u}_1$ and $\mathbf{u}_2$, with $AR = \varphi_L/\varphi_R \leq 1$, is defined by the head-balance form of Eq. (6) with the following internal description of the porosity discontinuity:

D1) the porosity varies monotonically between $\varphi_L$ and $\varphi_R$;

D2) the variation of flow depth and velocity through the porosity discontinuity is defined by a weak solution of Eq. (9) in the interval $s \in [0,1]$, with $\mathbf{v}(0) = \mathbf{u}_1$ and $\mathbf{v}(1) = \mathbf{u}_2$;

D3) the state $\mathbf{u}_2$ with $u_2 < 0$ is supercritical only if $|F(\mathbf{u}_2)| > K^*(AR)$; in this case, the relationship between $\mathbf{u}_1$ and $\mathbf{u}_2$ is defined by Eq. (6) where $\Delta H(\varphi_L, \varphi_R, \mathbf{u}_1, \mathbf{u}_2)$ is defined by Eq. (13).

To demonstrate the viability of the Definition 2, the exact solution to Riemann problems 7 and 8 of Table 1 is now found using the novel generalized Rankine-Hugoniot conditions. The 1-d exact solutions are represented with a black line in Figure 17, where the corresponding 2-d solution is represented with dots. The comparison with Figure 10, where the exact solutions are obtained with null energy loss, shows that introducing an appropriate definition of $\Delta H(\varphi_L, \varphi_R, \mathbf{u}_1, \mathbf{u}_2)$ reduces the discrepancy between the exact 1-d SP exact solution and the corresponding 2-d SWE numerical solution.

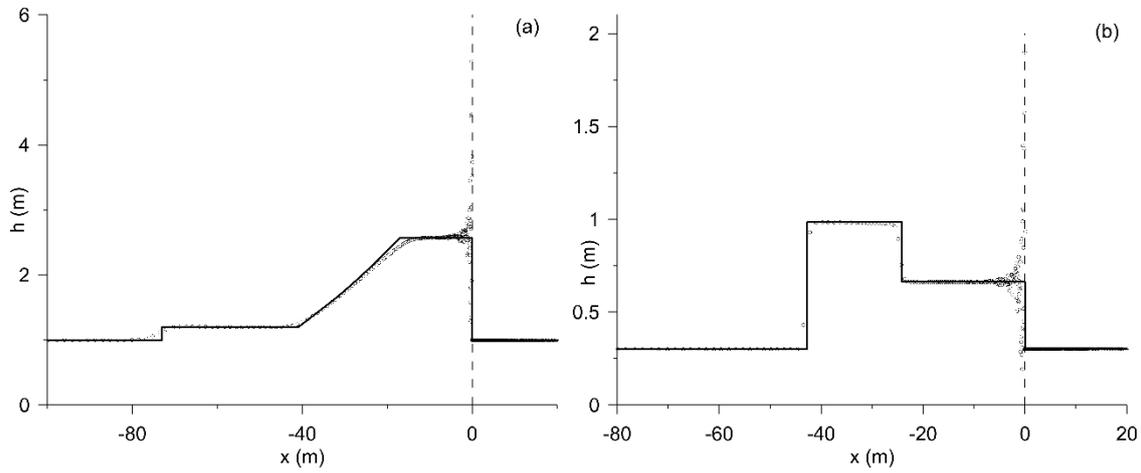

**Figure 17**. Profile view of the 1-d SP exact solution with head loss through the geometric discontinuity (continuous black line) and 2-d SWE numerical solution (dots) for the flow depth at time $t = 5$ s. Example Riemann problems 7 (a) and 8 (b) with initial conditions in Table 1.

## 5. Numerical model

In the present Section, the solution of the 1-d SP system of Eq. (2), where the initial conditions $\mathbf{u}(x,0) = \mathbf{u}_0(x)$ and the porosity distribution $\varphi(x)$ are specified, is approximated by means of the Finite

Volume method. Having partitioned the 1-d physical domain into non-overlapping cells $C_i = [x_{i-1/2}, x_{i+1/2}]$ of uniform length $\Delta x = x_{i+1/2} - x_{i-1/2}$, we assume that the averaged quantities

$$(14) \quad \varphi_i = \frac{1}{\Delta x} \int_{C_i} \varphi(x) dx, \quad \mathbf{u}_i^n = \left( h_i^n \quad h_i^n u_i^n \right)^T = \frac{1}{\varphi_i \Delta x} \int_{C_i} I(x) \mathbf{u}(x, t^n) dx$$

are approximations in $C_i$ of $\varphi(x)$ and $\mathbf{u}(x, t^n)$, respectively, where $t^n = n\Delta t$ is the time level. In Figure 18a, the cell-averaged constant values of $\varphi(x)$ and $\mathbf{u}(x, t^n)$ are conceptually depicted, showing the geometric and flow discontinuity at cells interface.

If $\Delta t = t_i^{n+1} - t_i^n$ is the time step length, the solution is advanced in the generic cell by means of the following explicit first-order scheme

$$(15) \quad \mathbf{u}_i^{n+1} = \mathbf{u}_i^n - \frac{\Delta t}{\varphi_i \Delta x} \left[ \psi_{i+1/2} \mathbf{g}\left(\mathbf{u}_{i+1/2}^-, \mathbf{u}_{i+1/2}^+\right) - \psi_{i-1/2} \mathbf{g}\left(\mathbf{u}_{i-1/2}^-, \mathbf{u}_{i-1/2}^+\right) \right] + \frac{\Delta t}{\varphi_i \Delta x} \left[ \mathbf{s}_{i-1/2}^+ + \mathbf{s}_{i+1/2}^- \right]$$

where the symbols are defined as follows: $\psi_{i+1/2}$ is a numerical approximation of the porosity at the interface $i+1/2$ between $C_i$ and $C_{i+1}$; $\mathbf{g}(\mathbf{u}, \mathbf{v})$ is a numerical flux corresponding to the 1-d SWE model in a constant width channel; finally, $\mathbf{u}_{i+1/2}^- = \left( h_{i+1/2}^- \quad h_{i+1/2}^- u_{i+1/2}^- \right)^T$ and $\mathbf{u}_{i+1/2}^+ = \left( h_{i+1/2}^+ \quad h_{i+1/2}^+ u_{i+1/2}^+ \right)^T$ are flow variables reconstructed to the left and right of the interface $i+1/2$, respectively, which are involved in the computation of numerical fluxes and non-conservative product approximations. The quantities $\mathbf{s}_{i-1/2}^+ = \left( 0 \quad s_{i-1/2}^+ \right)^T$ and $\mathbf{s}_{i+1/2}^- = \left( 0 \quad s_{i+1/2}^- \right)^T$ in Eq. (15) are the contributions to $C_i$ of the non-conservative products arising from the porosity gradient through the interfaces in $x_{i-1/2}$ and $x_{i+1/2}$, respectively, and their computation depends on the variable reconstruction adopted. The 1-d SWE numerical flux $\mathbf{g}(\mathbf{u},\mathbf{v})$ is approximated here by means of the HLLE Riemann solver (Cozzolino et al. 2014), although a different exact or approximate SWE Riemann solver could be used as well.

In its essence, the numerical scheme above consists of the following procedure. First, the porosity and flow interface variables are reconstructed from the cell-averaged values. Second, the numerical fluxes and porosity gradients contributions at interfaces are computed. Finally, the cell-averaged variables are advanced in time by means of Eq. (15). Clearly, the algorithm used to calculate the reconstructed interface variables from the cell-averaged values determines the properties of the scheme.

In the following, we assume without loss of generality that $\varphi_i \leq \varphi_{i+1}$, corresponding to $AR = \varphi_i/\varphi_{i+1} \leq 1$ (the procedure for the case $\varphi_i > \varphi_{i+1}$ is easily obtained by mirroring the local reference framework). The basic variable reconstruction by Castro et al. (2007) is first recalled, and then it is modified to introduce head losses through porosity discontinuities and cope with the case of multiple solutions. Finally, the results of the 1-d SP numerical scheme, with both the basic and novel interface variable reconstructions, are compared with the corresponding 2-d SWE numerical results.

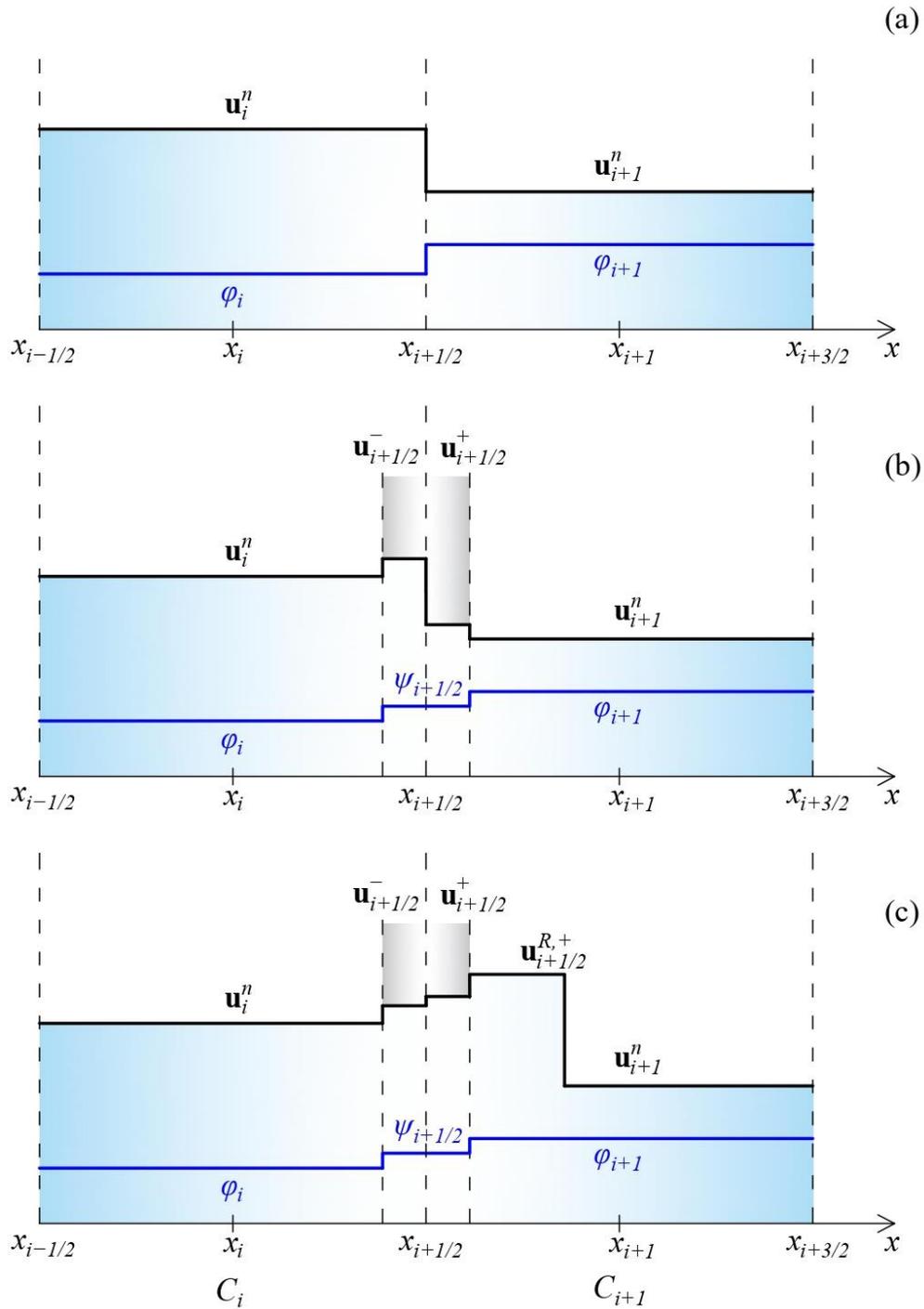

**Figure 18**. Side view of two neighbouring cells in the 1-d computational domain: cell-averaged quantities at a generic time level $n$ (a); interface reconstructed variables used in the basic reconstruction by Castro et al. (2007) (b); interface and in-cell reconstructed variables in the novel reconstruction approach (c).

*5.1 Basic reconstruction (Castro et al. 2007)*

The well-balanced reconstruction by Castro et al. (2007), originally implemented for the 1-d variable-width SWE model, is intended to capture steady state solutions where the discharge and head are uniform through the space domain. Aiming at this, the interface variables $\mathbf{u}^{-}_{i+1/2}$, $\mathbf{u}^{+}_{i+1/2}$, and $\psi_{i+1/2}$, are connected to the cell-averaged variables by means of Eq. (10), keeping the character of the flow (subcritical or supercritical). The reconstruction approach is schematically depicted in Figure 18b, where the porosity discontinuity internal structure is zoomed in to show the variables used for computations.

Given the right state $\mathbf{u}^{n}_{i+1}$, the following inequalities are checked (see Appendix C):

$$(16) \quad \left|F\left(\mathbf{u}^{n}_{i+1}\right)\right| < K_{sb}(AR), \quad \left|F\left(\mathbf{u}^{n}_{i+1}\right)\right| > K_{sp}(AR).$$

With reference to these checks, two options are possible, as follows.

*CR.1*) If one of the two inequalities in Eq. (16) is satisfied, the right state $\mathbf{u}^{n}_{i+1}$ can be connected to a state on the interface left side by the conditions of discharge and head invariance (see Section 2.2.2). In this case, the interface porosity $\psi_{i+1/2} = \varphi_i$ is assumed, and the state $\mathbf{u}^{+}_{i+1/2} = \left(h^{+}_{i+1/2} \quad h^{+}_{i+1/2} u^{+}_{i+1/2}\right)^{T}$ is easily found by solving the system

$$(17) \quad \begin{array}{l} \varphi_{i+1} h^{n}_{i+1} u^{n}_{i+1} - \psi_{i+1/2} h^{+}_{i+1/2} u^{+}_{i+1/2} = 0 \\ H\left(\mathbf{u}^{n}_{i+1}\right) - H\left(\mathbf{u}^{+}_{i+1/2}\right) = 0 \end{array},$$

which is obtained by assuming in Eq. (10) the positions $\varphi_L = \psi_{i+1/2}$, $\varphi_R = \varphi_{i+1}$, $\mathbf{u}_1 = \mathbf{u}^{+}_{i+1/2}$, and $\mathbf{u}_2 = \mathbf{u}^{n}_{i+1}$. The system of Eq. (17) admits two exact solutions, one corresponding to a subcritical state and the other to a supercritical state (see Valiani and Caleffi 2008 for the corresponding exact expressions). The first is chosen if the state $\mathbf{u}_{i+1}^n$ is subcritical, otherwise the supercritical one is kept. Finally, the position $\mathbf{u}_{i+1/2}^- = \mathbf{u}_i^n$ is made.

*CR.2*) If the inequalities of Eq. (16) are not satisfied, the system of Eq. (17) admits no solution with $\psi_{i+1/2} = \varphi_i$. In this case, $\mathbf{u}_{i+1/2}^+$ is found by means of Eq. (17) where the interface porosity $\psi_{i+1/2}$ is defined as

$$(18) \quad \psi_{i+1/2} = \varphi_{i+1} \left| F\left(\mathbf{u}_{i+1}^n\right) \right| \left( \frac{3}{2 + F^2\left(\mathbf{u}_{i+1}^n\right)} \right)^{3/2}.$$

This choice is equivalent to imposing that $\mathbf{u}_{i+1/2}^+$ is critical (see Appendix C). The state $\mathbf{u}_{i+1/2}^-$ is calculated by means of

$$(19) \quad \begin{array}{l} \psi_{i+1/2} h_{i+1/2}^- u_{i+1/2}^- - \varphi_i h_i^n u_i^n = 0 \\ H\left(\mathbf{u}_{i+1/2}^-\right) - H\left(\mathbf{u}_i^n\right) = 0 \end{array},$$

which is obtained by assuming in Eq. (10) the positions $\varphi_L = \varphi_i$, $\varphi_R = \psi_{i+1/2}$, $\mathbf{u}_1 = \mathbf{u}_i^n$, and $\mathbf{u}_2 = \mathbf{u}_{i+1/2}^-$. The subcritical solution is kept if the state $\mathbf{u}_i^n$ is subcritical, otherwise the supercritical solution is chosen. The state $\mathbf{u}_{i+1/2}^-$ certainly exists because $\varphi_i < \psi_{i+1/2}$ (see Appendix C).

From the preceding, it is evident that the reconstruction by Castro et al. (2007) satisfies the inequality $\min\{\varphi_i, \varphi_{i+1}\} \leq \psi_{i+1/2} \leq \max\{\varphi_i, \varphi_{i+1}\}$, i.e., it ensures the monotonicity of the porosity variation through the discontinuity.

The algorithm is completed by using Eqs. (5.a)-(5.b) to express $\mathbf{s}^+_{i-1/2}$ and $\mathbf{s}^-_{i+1/2}$ as

$$
\begin{aligned}
\mathbf{s}^+_{i-1/2} &= \varphi_i \mathbf{f}\left(\mathbf{u}^n_i\right) - \psi_{i-1/2} \mathbf{f}\left(\mathbf{u}^+_{i-1/2}\right) \\
\mathbf{s}^-_{i+1/2} &= \psi_{i+1/2} \mathbf{f}\left(\mathbf{u}^-_{i+1/2}\right) - \varphi_i \mathbf{f}\left(\mathbf{u}^n_i\right)
\end{aligned}
\qquad (20)
$$

*5.2 Novel variable reconstruction*

The novel variable reconstruction differs from that by Castro et al. (2007) because the case of a supercritical flow impinging on a porosity reduction is treated in a separate way, congruently with the novel definition of Rankine-Hugoniot conditions given in Section 4.3. This is accomplished by computing an in-cell additional reconstructed state $\mathbf{u}^{R,+}_{i+1/2} = \left(h^{R,+}_{i+1/2} \quad h^{R,+}_{i+1/2} u^{R,+}_{i+1/2}\right)^T$ to manage the case of a backwards moving shock between the geometric transition and the state $\mathbf{u}^n_{i+1}$ when a T3 solution (Figure 4c) occurs. In addition, appropriate head loss is introduced to compute the state $\mathbf{u}^+_{i+1/2}$ if the flow through the porosity reduction is supercritical. The novel reconstruction approach is schematically depicted in Figure 18c, showing the in-cell additional reconstructed variable $\mathbf{u}^{R,+}_{i+1/2}$.

Given the right state $\mathbf{u}^n_{i+1}$, the cases $u^n_{i+1} < 0$ and $u^n_{i+1} \geq 0$ are treated separately.

*NR.1)* If $u^n_{i+1} \geq 0$, the procedure by Castro et al. (2007) is used to find $\psi_{i+1/2}$, $\mathbf{u}^-_{i+1/2}$, and $\mathbf{u}^+_{i+1/2}$ (see points CR.1 and CR.2 of Section 5.1). In this case, there is no backwards moving shock and the position $\mathbf{u}^{R,+}_{i+1/2} = \mathbf{u}^n_{i+1}$ is made.

*NR.2)* If $u_{i+1}^n < 0$, the quantity $\left|F\left(\mathbf{u}_{i+1}^n\right)\right|$ is compared to $K_{sb}(AR)$ and $K_{jump}^*(AR)$. Three cases are possible:

*NR.2.1)* If $\left|F\left(\mathbf{u}_{i+1}^n\right)\right| < K_{sb}(AR)$ occurs, the energy of the subcritical flow is sufficient to pass through the porosity reduction, and the point CR.1 of Section 5.1 supplies $\psi_{i+1/2}$, $\mathbf{u}_{i+1/2}^-$, and $\mathbf{u}_{i+1/2}^+$. The position $\mathbf{u}_{i+1/2}^{R,+} = \mathbf{u}_{i+1}^n$ is made because there is no backwards moving shock.

*NR.2.2)* If $\left|F\left(\mathbf{u}_{i+1}^n\right)\right| > K_{jump}^*(AR)$, the energy of the supercritical flow is sufficient to pass through the porosity reduction with head losses (see Section 4.2). In this case, the interface porosity $\psi_{i+1/2} = \varphi_i$ is assumed and the state $\mathbf{u}_{i+1/2}^+$ is easily found by picking the supercritical solution of the system

$$(21) \quad \begin{aligned} \varphi_{i+1} h_{i+1}^n u_{i+1}^n - \psi_{i+1/2} h_{i+1/2}^+ u_{i+1/2}^+ = 0 \\ H\left(\mathbf{u}_{i+1}^n\right) - H\left(\mathbf{u}_{i+1/2}^+\right) = \Delta H\left(\psi_{i+1/2}, \varphi_{i+1}, \mathbf{u}_{i+1/2}^+, \mathbf{u}_{i+1}^n\right) \end{aligned},$$

which is obtained by assuming in Eq. (10) the positions $\varphi_L = \psi_{i+1/2}$, $\varphi_R = \varphi_{i+1}$, $\mathbf{u}_1 = \mathbf{u}_{i+1/2}^+$, and $\mathbf{u}_2 = \mathbf{u}_{i+1}^n$. The head loss in Eq. (21) is computed using the Eqs. (12) and (13). Finally, the positions $\mathbf{u}_{i+1/2}^- = \mathbf{u}_i^n$ and $\mathbf{u}_{i+1/2}^{R,+} = \mathbf{u}_{i+1}^n$ are made.

*NR.2.3)* If $K_{sb}(AR) \leq \left|F\left(\mathbf{u}_{i+1}^n\right)\right| \leq K_{jump}^*(AR)$, the energy of the state $\mathbf{u}_{i+1}^n$ is either insufficient to pass through the porosity reduction or a multiple solution is possible. In both the cases, the Riemann problem solution is characterized by a backwards moving shock radiating from the geometric discontinuity. The occurrence of this shock is forced by posing $\psi_{i+1/2} = \varphi_i$ and assuming that the states $\mathbf{u}^+_{i+1/2}$ and $\mathbf{u}^{R,+}_{i+1/2}$ have the same discharge of the state $\mathbf{u}^n_{i+1}$. The state $\mathbf{u}^+_{i+1/2}$ is critical, obtaining

$$(22) \quad \begin{aligned} &\varphi_{i+1} h^n_{i+1} u^n_{i+1} - \psi_{i+1/2} h^+_{i+1/2} u^+_{i+1/2} = 0 \\ &u^+_{i+1/2} = -\sqrt{g h^+_{i+1/2}} \end{aligned},$$

while the state $\mathbf{u}^{R,+}_{i+1/2}$ is subcritical with $F\left(\mathbf{u}^{R,+}_{i+1/2}\right) = -K^*_{jump}(AR)$, obtaining

$$(23) \quad \begin{aligned} &h^n_{i+1} u^n_{i+1} - h^{R,+}_{i+1/2} u^{R,+}_{i+1/2} = 0 \\ &u^{R,+}_{i+1/2} = -\sqrt{g h^{R,+}_{i+1/2}} K^*_{jump}(AR) \end{aligned}.$$

Finally, the position $\mathbf{u}^-_{i+1/2} = \mathbf{u}^n_i$ is made.

The novel reconstruction satisfies the inequality $\min\{\varphi_i, \varphi_{i+1}\} \leq \psi_{i+1/2} \leq \max\{\varphi_i, \varphi_{i+1}\}$, ensuring the monotonicity of the porosity discontinuity inner description. Having introduced the in-cell additional reconstructed state $\mathbf{u}^{R,+}_{i+1/2}$, the definition of $\mathbf{s}^+_{i-1/2}$ in Eq. (20) changes as follows:

$$(24) \quad \mathbf{s}^+_{i-1/2} = \varphi_i \mathbf{f}\left(\mathbf{u}^{R,+}_{i-1/2}\right) - \psi_{i-1/2} \mathbf{f}\left(\mathbf{u}^+_{i-1/2}\right).$$

A similar reconstruction approach has been proposed by Varra et al. (2022), where an iterative procedure is used. Nonetheless, the present reconstruction is an improvement because (i) the iterative procedure is avoided and (ii) it is possible to consider the cases where $K^*_{jump}(AR) > K_{sp}(AR)$. In addition, adequate head loss is introduced for supercritical flows through abrupt porosity reductions.

*5.3 Numerical experiments*

The 1-d numerical model of Eq. (15), equipped with the variable reconstructions described in Sections 5.1 and 5.2, respectively, is used to approximate the solution of the Riemann problems with initial conditions in Table 1. For the sake of simplicity, $\Delta x = 0.20$ m and $\Delta t = 0.005$ s in all the numerical experiments.

*5.3.1 Numerical experiments with the reconstruction by Castro et al. (2007)*

Figure 19 shows the numerical results (flow depth) to the Riemann problems 1-4 supplied by the 1-d SP model with the reconstruction by Castro et al. (2007) (continuous black line), while the corresponding 2-d SWE results are represented with white dots. For all these problems, the algorithm by Castro et al. (2007) captures the essentials of the 2-d SWE solution, namely the number of waves and their strength, and the flow depth of the intermediate states. The discrepancies between the 1-d and 2-d solutions in Figures 19b and 19c (Riemann problems 2 and 3, respectively) correspond to those discussed with reference to the comparison between Riemann exact solution and 2-d numerical solution (compare with Figures 6c and 6d).

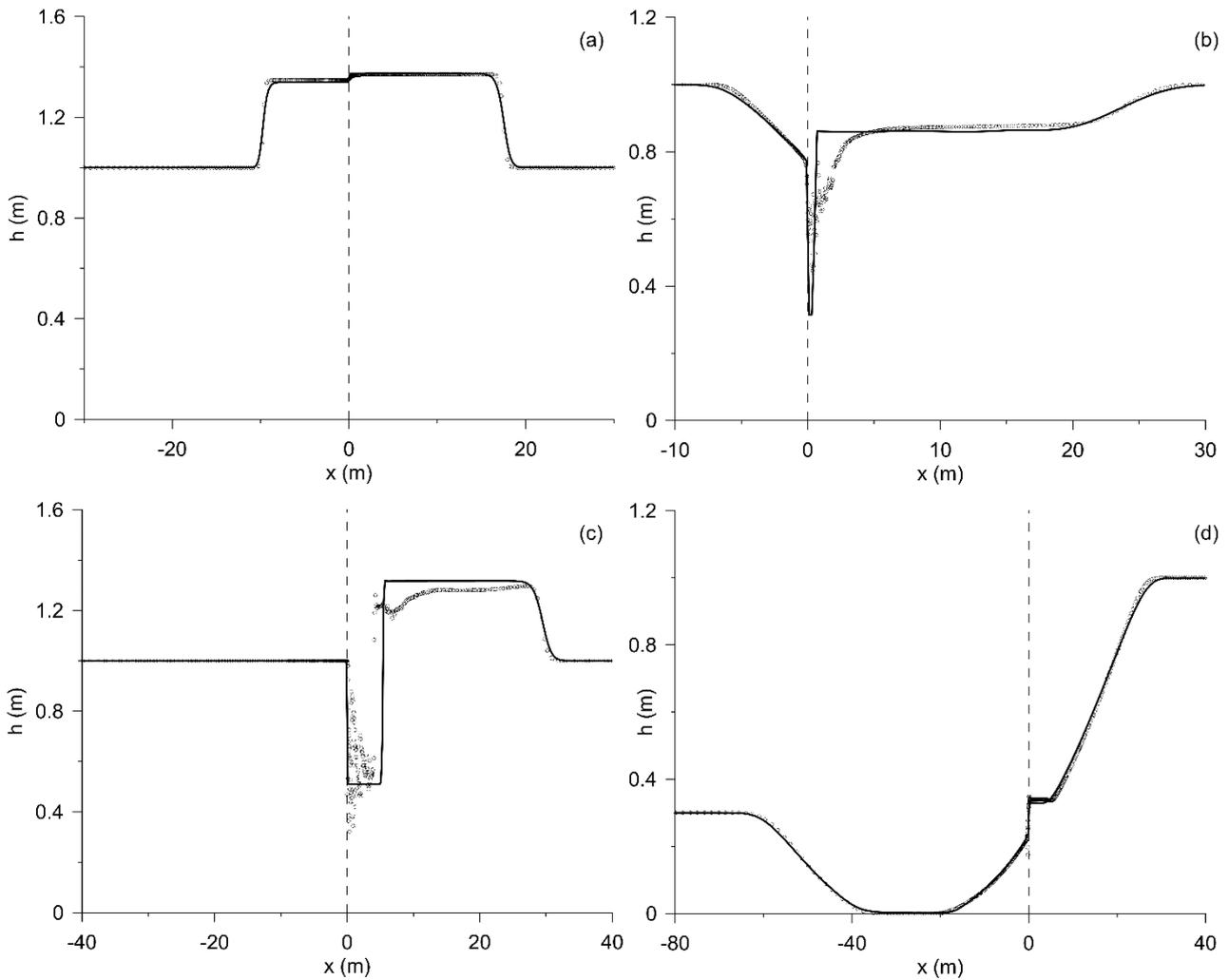

**Figure 19**. Profile view of the numerical solution for the flow depth at time $t = 5$ s: 1-d SP model with the variable reconstruction by Castro et al. (2007) (continuous black line) and 2-d SWE model (dots). Example Riemann problems 1 (a), 2 (b), 3 (c) and 4 (d) with initial conditions in Table 1.

The numerical results to Riemann problems 5-8 are represented in Figure 20. These cases, characterized by a supercritical flow through a porosity reduction, show that the 1-d SP numerical solution with the basic reconstruction by Castro et al. (2007) greatly differs from the corresponding 2-d SWE reference solution.

The Figures 20a,b refer to Riemann problems (5 and 6, respectively) admitting multiple solutions. While the 2-d SWE model exhibits a backwards moving shock originated from the porosity discontinuity that causes flow energy dissipation, the 1-d SP numerical model with variable reconstruction by Castro et al. (2007) captures the solution with supercritical flow through the discontinuity. This is expected because the reconstruction by Castro et al. (2007) keeps the supercritical character of the flow impinging on the porosity reduction (point CR.1 in Section 5.1). From the preceding, it follows that the numerical scheme by Castro et al. (2007) overestimates the discharge through the geometric discontinuity and the celerity of the advancing shock on the left, while completely neglects the energy dissipation mechanism connected with the backwards moving shock generated by the interaction of the propagating flow with obstacles (Guinot et al. 2017). Varra et al. (2020) have demonstrated that the same defect is shared by other Riemann solvers such as those by Cozzolino et al. (2018b) and Guinot et al. (2017).

The Figures 20c,d refer to Riemann problems (7 and 8, respectively) where a supercritical flow impinges on a porosity reduction, but the solution is unique. In these cases, the 1-d SP numerical model misses to capture the 2-d SWE model solution because it lacks an appropriate dissipation mechanism. Interestingly, this is the same discrepancy found when comparing the 1-d exact solution and the 2-d SWE numerical solution (see Figures 10a,b).

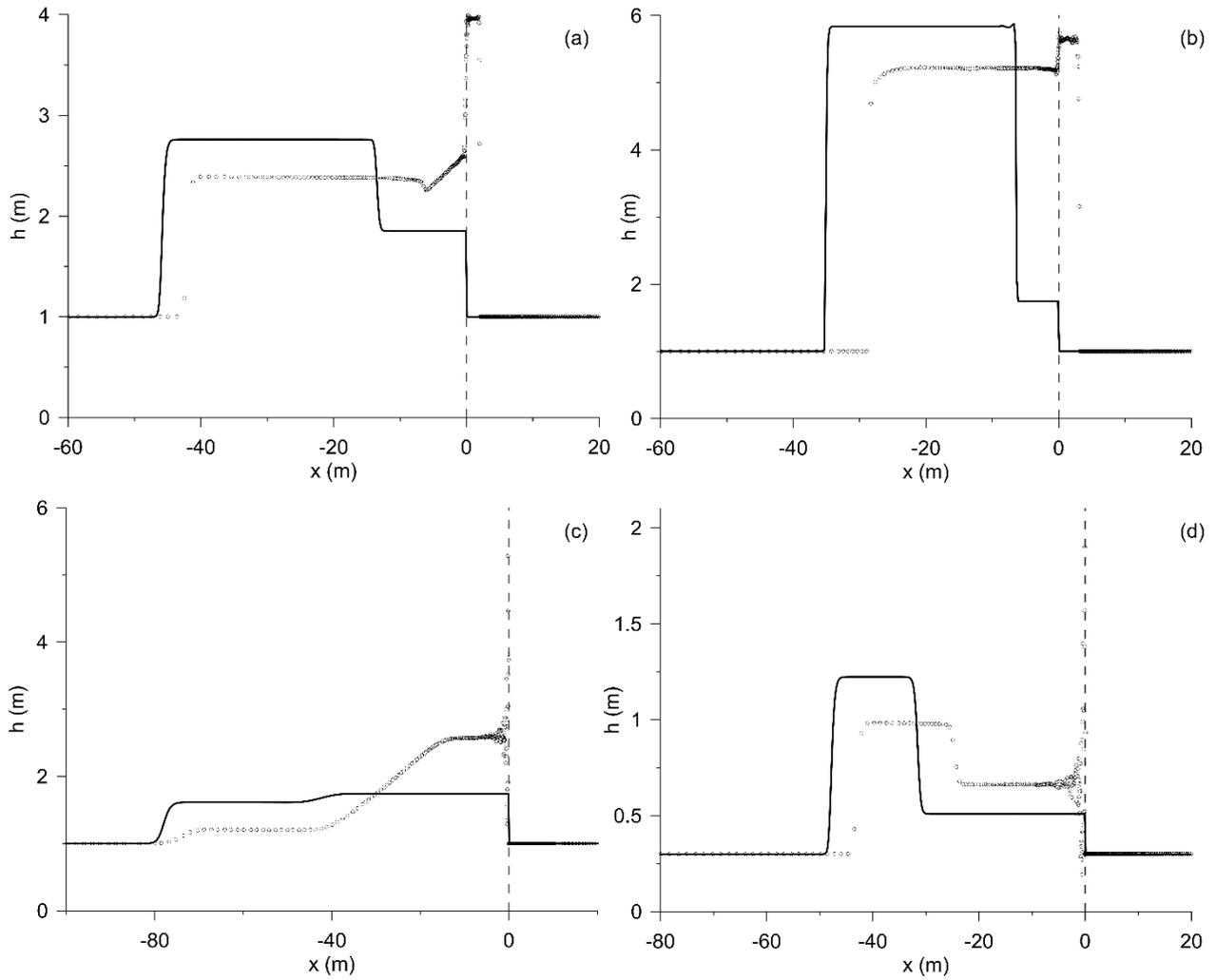

**Figure 20**. Profile view of the numerical solution for the flow depth at time $t = 5$ s: 1-d SP model with the basic variable reconstruction by Castro et al. (2007) (continuous black line) and 2-d SWE model (dots). Example Riemann problems 5 (a), 6 (b), 7 (c) and 8 (d) with initial conditions in Table 1.

*5.3.2 Numerical experiments with the novel reconstruction*

The numerical experiments presented in the preceding subsection are repeated using the 1-d SP numerical model with with the novel reconstruction of Section 5.2. For the Riemann problems 1-4, the numerical results supplied by the novel reconstruction, which are not reported here for the sake of brevity, coincide with those supplied by the basic reconstruction by Castro et al. (2007). This is expected, because these Riemann problems do not refer to cases where a supercritical flow impinges on a porosity reduction.

The Figures 21a,b refer to the Riemann problems 5 and 6, respectively, which admit multiple solutions. Contrary to the algorithm by Castro et al. (2007), the novel variable reconstruction captures the solution with the backwards moving shock exhibited by the 2-d SWE model. It is clear that the introduction of the in-cell subcritical state $\mathbf{u}_{i+1/2}^{R,+}$ between the geometric transition and the right state $\mathbf{u}_i^n$, together with the computation of the interface reaction term by means of Eq. (24), is the ingredient allowing the computation of the physically congruent shock immediately to the discontinuity right-side.

The comparison between the 1-d SP numerical results obtained with the novel variable reconstruction and the 2-d SWE results for Riemann problems 7 and 8, is represented in Figures 21c,d, respectively. The inspection of these figures, referring to cases where the supercritical flow impinging on the porosity reduction remains supercritical with loss of energy, shows that the novel variable reconstruction satisfactorily reproduces the 2-d SWE model results because the required amount of head loss through the geometric discontinuity is introduced.

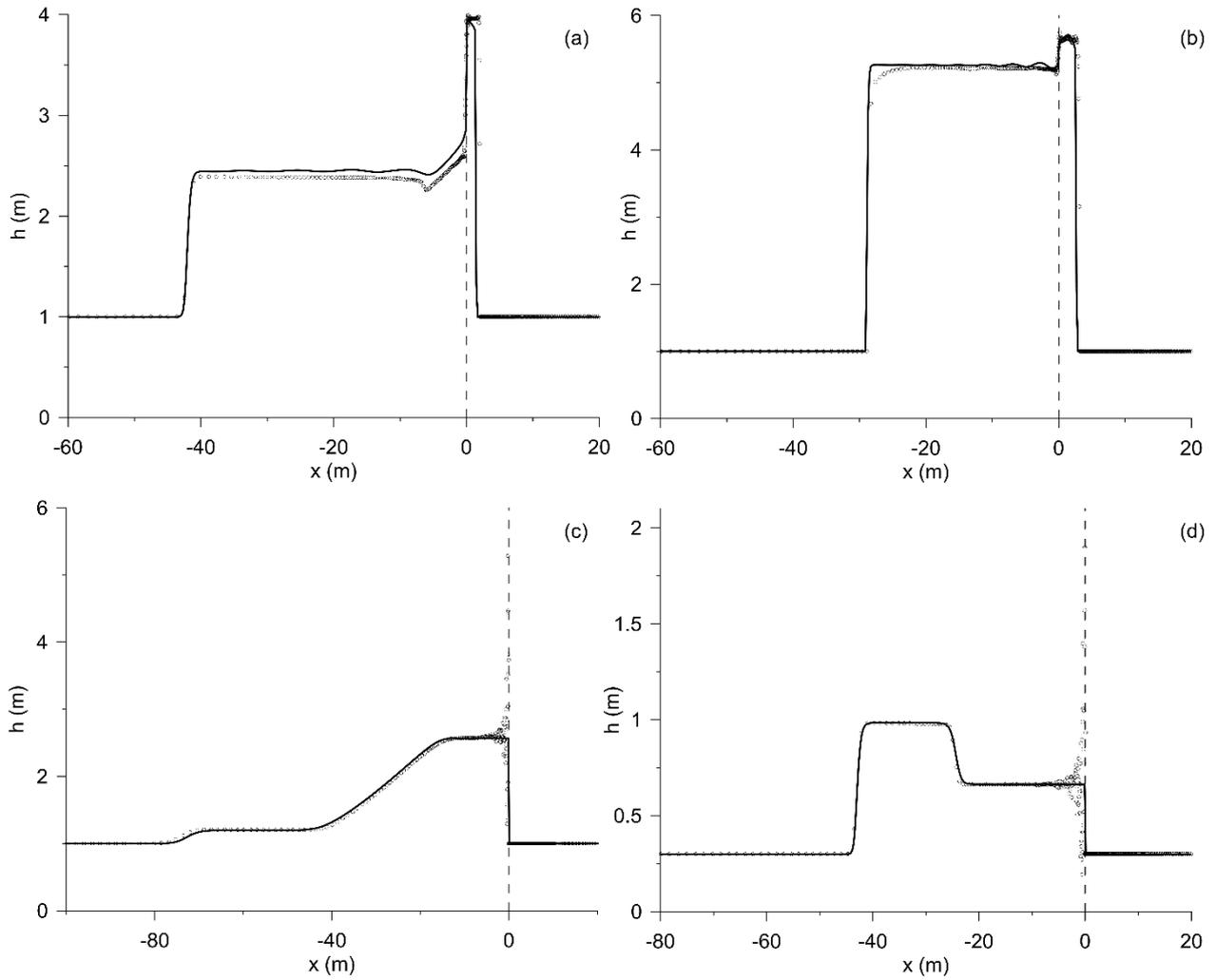

**Figure 21**. Profile view of the numerical solution for the flow depth at time $t = 5$ s: 1-d SP model with the novel variable reconstruction (continuous black line) and 2-d SWE model (dots). Example Riemann problems 5 (a), 6 (b), 7 (c) and 8 (d) with initial conditions in Table 1.

## 6. Discussion

In this section, the model presented in the preceding sections is discussed with reference to alternative numerical and conceptual approaches available in the literature, and with reference to the robustness of Eqs. (11) and (12).

*6.1 Comparison with the transient momentum dissipation approach*

The momentum dissipation approach introduced with the DIP numerical model by Guinot et al. (2017) is a numerical device intended to introduce the transient energy dissipation generated by bore reflection at obstacles during transient propagation. In the 1-d case, the DIP model can be written as

$$(25) \quad \mathbf{u}_i^{n+1} = \mathbf{u}_i^n - \frac{\Delta t}{\varphi_i \Delta x}\left[\psi_{i+1/2}\mathbf{M}_{i+1/2}\mathbf{g}\left(\mathbf{u}_{i+1/2}^-, \mathbf{u}_{i+1/2}^+\right) - \psi_{i-1/2}\mathbf{M}_{i-1/2}\mathbf{g}\left(\mathbf{u}_{i-1/2}^-, \mathbf{u}_{i-1/2}^+\right)\right] + \frac{\Delta t}{\varphi_i \Delta x}\left[\mathbf{s}_{sta,i-1/2}^+ + \mathbf{s}_{sta,i+1/2}^-\right]$$

,

where **M** is a momentum dissipation matrix defined as (Guinot et al. 2017)

$$(26) \quad \mathbf{M} = \begin{pmatrix} 1 & 0 \\ 0 & 1-\mu \end{pmatrix},$$

with $\mu > 0$ in case $h_i^{n+1} > h_i^n$, while $\mu = 0$ otherwise. The momentum dissipation coefficient $\mu$ appearing in the matrix **M** of Eq. (26) must be calibrated using fine grid 2-d SWE simulations and it is strongly dependent on the urban fabric structure and flow conditions (Guinot et al. 2017).

In Eq. (25), the interface porosity $\psi_{i+1/2}$ is evaluated from the underlying urban fabric with a non-monotonic approach, which enforces the condition $\psi_{i+1/2} \leq \min(\varphi_i, \varphi_{i+1})$ (see Section 2.2.1); the interface contributions $\mathbf{s}_{sta,i-1/2}^+$ and $\mathbf{s}_{sta,i+1/2}^-$ of the non-conservative products are computed under the assumption of in-cell stagnant water (Guinot and Soares-Frazão 2006) as

$$(27) \quad \begin{aligned} \mathbf{s}_{sta,i-1/2}^+ &= \left(\varphi_i - \psi_{i-1/2}\right)\frac{g}{2}\left(h_i^n\right)^2 (0 \ \ 1)^T \\ \mathbf{s}_{sta,i+1/2}^- &= \left(\psi_{i+1/2} - \varphi_i\right)\frac{g}{2}\left(h_i^n\right)^2 (0 \ \ 1)^T \end{aligned}.$$

Finally, the reconstructed variables are defined as:

$$(28) \quad \begin{aligned} \mathbf{u}_{i+1/2}^{-} &= \left( h_i^n \quad \varphi_i h_i^n u_i^n / \psi_{i+1/2} \right)^T \\ \mathbf{u}_{i+1/2}^{+} &= \left( h_{i+1}^n \quad \varphi_{i+1} h_{i+1}^n u_{i+1}^n / \psi_{i+1/2} \right)^T \end{aligned}.$$

We reinterpret the momentum dissipation approach observing that the 1-d DIP numerical model of Eq. (25) can be rewritten in the form of Eq. (15) by defining the interface contributions of the non-conservative products as

$$(29) \quad \begin{aligned} \mathbf{s}_{i-1/2}^{+} &= \mathbf{s}_{sta,i-1/2}^{+} + \psi_{i-1/2} \left( \mathbf{M}_{i-1/2} - \mathbf{I} \right) \mathbf{g} \left( \mathbf{u}_{i-1/2}^{-}, \mathbf{u}_{i-1/2}^{+} \right) \\ \mathbf{s}_{i+1/2}^{-} &= \mathbf{s}_{sta,i+1/2}^{-} + \psi_{i+1/2} \left( \mathbf{I} - \mathbf{M}_{i+1/2} \right) \mathbf{g} \left( \mathbf{u}_{i+1/2}^{-}, \mathbf{u}_{i+1/2}^{+} \right) \end{aligned}.$$

According to this reinterpretation, the momentum dissipation approach is equivalent to evaluating the forces exerted by obstacles at cell interfaces by adding or subtracting a dynamic contribution to the stagnant water hydrostatic thrusts $\mathbf{s}_{sta,i-1/2}^{+}$ and $\mathbf{s}_{sta,i+1/2}^{-}$. A slightly different definition of the matrix $\mathbf{M}$ has been subsequently given in Guinot et al. (2018), but the role of $\mathbf{M}$ as a modulator of the forces exerted by obstacles at cell interfaces remains unchanged.

The reinterpretation supplied by Eq. (29) allows to recognize the common numerical framework of the DIP model (Guinot et al. 2017) and of the numerical model presented here. Nonetheless, the exact solution of the 1-d SP Riemann problem has been exploited in the present paper to introduce a mechanism of energy dissipation caused by the reflection of advancing waves at porosity discontinuities. This approach, which has been accomplished by isolating a single local porosity discontinuity and comparing the corresponding 1-d SP and 2-d SWE Riemann solutions (see Section 3), avoids the intricacies caused by the mutual interaction of waves radiating from the obstacles of complex urban fabrics and allows to consider the local geometry. In addition, it avoids the introduction of a momentum dissipation mechanism of unclear physical meaning, whose parameters need calibration on a case-by-case basis (Guinot et al. 2017).

*6.2 Disambiguation of the porosity Riemann problem*

The issue of disambiguating multiple solutions to the Riemann problem where a geometric discontinuity is present has been tackled by researchers considering different types of geometric discontinuities or different fluid models (SWE or Euler equations). It is interesting to compare the results obtained in the present paper with those available in the literature.

The exact solution to the Riemann problem for the 1-d SWE model with variable bed elevation exhibits two classes of triple solutions (for convenience, the solutions are called here S1, S2, and S3) when a supercritical flow impinges on a positive bed step (Han and Warnecke 2014, Aleksyuk et al. 2022). In the first class of triple solutions, S1 is characterised by a supercritical flow that jumps over the bed step remaining supercritical, while S2 is characterised by a hydraulic jump located through the discontinuity that reverts the incoming supercritical flow into subcritical. Finally, S3 is characterised by a backward shock while subcritical flow conditions are established over the bed step. The second class of triple solutions differs from the first one because the solution S3 is characterised by a backward shock with blockage of the flow at the bed step (step higher than the free surface level). Cozzolino et al. (2014) used a mix of steady state laboratory data (Karki et al. 1972, Hager and Sinniger 1985) and physical reasoning to establish a disambiguation criterion based on discharge minimization. Following this criterion, the physically relevant solution is S3 (backward shock) in both the classes of triple solutions. Aleksyuk and Belikov (2019) found the same result by considering a mathematical argument based on the continuous dependence of solutions on the initial conditions.

Han et al. (2013) considered the Riemann problem for the 1-d Euler equations in a compressible duct flow, where triple solutions may occur when a supersonic flow impinges on a pipe diameter reduction. They compared some examples of 1-d exact multiple solutions with the numerical results supplied by a higher dimensions axisymmetric Euler equations model (longitudinal and radial direction), founding that the physically relevant solutions were those characterized by a backward shock.

A pattern seems to emerge from these results. In all the examples considered, multiple solutions occur when a supercritical flow (SWE model) or a supersonic flow (Euler equations) impact on a cross-section reduction. Despite the variety of mathematical models and means applied for the disambiguation of multiple exact solutions (laboratory and/or numerical experiments, mathematical arguments), the common output to the different procedures is that the physically relevant solution among the alternatives is the one characterised by a backward shock. In SWE models, this is also the solution which minimizes the discharge through the geometric discontinuity.

The results presented in Section 4, which confirm this pattern for the 1-d SP model, can be reinterpreted using an argument based on the continuous dependence of the Riemann problem solution on the initial conditions. In fact, when the tailwater is null ($h_L = 0$), the 1-d SP Riemann problem exhibits three exact solutions (T1, T2, and T3) in the region B of Figure 3, which is bounded by the curves LB and UB, and one exact solution in regions A and C (T3 and T1, respectively). A solution with backward shock (T3 of Figure 4c) in region B of Figure 3 can move through LB to a solution T3 of region A by decreasing the initial Froude number $|F_R|$. On the other side, the same T3 solution in region B can move through UB to a T1 solution of region C by increasing $|F_R|$, flushing the hydraulic jump and establishing supercritical flow conditions through the porosity discontinuity. In conclusion, T3 solutions should be considered to the left of UB, and T1 solutions to the right. Of course, the head loss generated by transverse shocks through the porosity discontinuity does not significantly changes this picture. In this case, the limit curve UB is distorted, becoming the modified limit curve MUB of Figure 14. Indeed, the initial conditions to the left of the MUB curve characterize solutions with a backward shock (G2 solutions in Section 4.1), while the initial conditions to the right characterize supercritical flows through the porosity discontinuity (G1 solutions).

## 6.3 Influence of flow depth and channel width on the porosity discontinuity definition

The numerical experiments of Section 4 have been conducted considering a fixed depth $h_R = 1$ m of the flow impinging on a channel contraction and fixed width $B_R = 1$ m of the right channel reach. The reader may wonder if these results can be extended to different values of $h_R$ and $B_R$. The answer to this question is affirmative but it requires a brief discussion.

Recalling that viscosity and density are not modelled by the incompressible SWE model, the most general way to write the expression of the head loss $\Delta H^*$ suffered by a supercritical flow $\mathbf{u}_R$ through the contraction is

(30) $\Delta H^* = f\left(B_L, B_R, h_R, u_R, g, L_c\right).$

We observe that $n = 7$ physical quantities are involved in Eq. (30), and we want to ascertain if the physical equation can be simplified. Aiming at this, we additionally see that only $k = 2$ fundamental mechanics units, namely length and time, are involved because there is not dependency on density. Recalling the Vaschy-Buckingham theorem (Vaschy 1892, Buckingham 1914), it follows that Eq. (30) can be rewritten as an expression involving $n - k = 5$ dimensionless independent quantities. The dimensionless quantities chosen are

(31) $AR = \dfrac{B_L}{B_R}, \quad F_R = \dfrac{u_R}{\sqrt{gh_R}}, \quad \Delta^* = \dfrac{\Delta H^*}{h_R + u_R^2/(2g)}, \quad \sigma_R = \dfrac{h_R}{B_R}, \quad \lambda = \dfrac{B_R - B_L}{L_c},$

and it is easy to verify that they are mutually independent, i.e., it is not possible to build one of the dimensionless quantities starting from the others. This justifies why it is possible to simplify Eq. (30) as

(32) $\Delta^* = f\left(AR, F_R^2, \sigma_R, \lambda\right).$

In a similar manner, a very general way to describe the boundary MUB between the G1 and G2 solutions discussed in Section 4 is to introduce a limit velocity $u_R^*$ discriminating the two types of solution and expressing this velocity as

(33) $u_R^* = f(B_L, B_R, h_R, g, L_c)$.

This physical equation involves $n = 6$ physical quantities and $k = 2$ fundamental units, implying that it can be simplified as

(34) $K^* = f(AR, \sigma_R, \lambda)$

where $K^* = |u_R^*|/\sqrt{gh_R}$.

We observe that the dependence on $\lambda$ does not need to be explicited because this parameter is constant in all the 2-d SWE simulations, since the contraction walls are always inclined by 45° with respect to the channel axis. With reference to the parameter $\sigma_R$, the comparison between Eq. (32) and (12), and between Eq. (34) and (11), respectively, show that the expressions found in Section 4 are valid in the case $\sigma_R = 1$ because $h_R = 1$ m and $B_R = 1$ m in all the numerical experiments. Nonetheless, we demonstrate that the parameter $\sigma_R$ is superfluous because the expressions of Eqs. (32) and (34) can be safely simplified in the form of Eqs. (12) and (11), respectively.

Consider the steady state 1-d SWE in a channel of variable width $B = B(x)$, with solution $\mathbf{u} = \mathbf{u}(x)$ for given right boundary condition $\mathbf{u}_R$:

(35) $\dfrac{dB\mathbf{f}(\mathbf{u})}{dx} + \mathbf{h}(\mathbf{u})\dfrac{dB}{dx} = 0$.

Despite Eq. (35) is a simplification of the 2-d flow through the contraction, it supplies a sufficient insight for the present discussion. We observe that the multiplication of Eq. (35) by the constant $k$ allows to write

$$(36) \quad k\left(\frac{dB\mathbf{f}(\mathbf{u})}{dx} + \mathbf{h}(\mathbf{u})\frac{dB}{dx}\right) = 0,$$

which is still satisfied by the solution $\mathbf{u} = \mathbf{u}(x)$ of Eq. (35). Of course, $k$ can be moved inside the derivative symbols, leading to

$$(37) \quad \frac{dB'\mathbf{f}(\mathbf{u})}{dx} + \mathbf{h}(\mathbf{u})\frac{dB'}{dx} = 0,$$

where $B'(x) = kB(x)$. In other words, the solution $\mathbf{u} = \mathbf{u}(x)$ of Eq. (35) for given boundary condition $\mathbf{u}_R$ does not change if the width is uniformly amplified by a constant $k$. The uniform amplification of the width affects the parameter $\sigma_R = h_R/B_R$ but does not affect the parameters $AR = B_L/B_R$ and $F_R = u_R/\sqrt{gh_R}$, implying that Eqs. (32) and (34) depend on $AR$ and $F_R^2$ but not on $\sigma_R$.

To evaluate the robustness of this theoretical approach, we consider twelve additional 2-d SWE simulations with initial and geometrical conditions as follows: $AR = 0.3, 0.6$; $|F_R| = 3.6, 6, 8, 11$; $h_R = 0.1, 0.5, 1$ m, and $B_R = 1$ m (see Table 4). The values chosen for $h_R$ correspond to the three different values ($\sigma_R = 0.1, 0.5,$ and $1$) of the dimensionless parameter $\sigma_R$. Six of the chosen flow conditions correspond to points slightly to the left of the MUB curve and six to the right (see Figure 22). Table 4 reports, for each simulation, the dimensionless head loss $\Delta^*$ if the solution is of type G1, while a hyphen indicates a G2-type solution. The inspection of these results shows that the solution types (G1 or G2) expected based on the position with respect to the MUB curve are those actually occurring in the 2-d simulations, while the dependence of the relative head loss $\Delta^*$ on $\sigma_R$ is negligible. These observations confirm the theoretical arguments above and justify the application of the formulas in Section 4 to conditions with $\sigma_R \neq 1$.

**Table 4**. Supercritical 2-d flow impinging on a contraction for $B_R = 1$ m and $h_R = 0.1, 0.5,$ and 1 m: relative head losses $\Delta^*$ for the cases of flow passing through the discontinuity (G1). A hyphen indicates the cases where a backwards moving shock is produced (G2 configurations).

|           | *AR*       |        |           |        |
|-----------|------------|--------|-----------|--------|
|           | 0.30       |        | 0.60      |        |
|           | $\|F_R\|$  |        | $\|F_R\|$ |        |
| $h_R$ (m) | 8          | 11     | 3.6       | 6      |
| 0.1       | -          | 0.57   | -         | 0.38   |
| 0.5       | -          | 0.57   | -         | 0.38   |
| 1         | -          | 0.57   | -         | 0.38   |

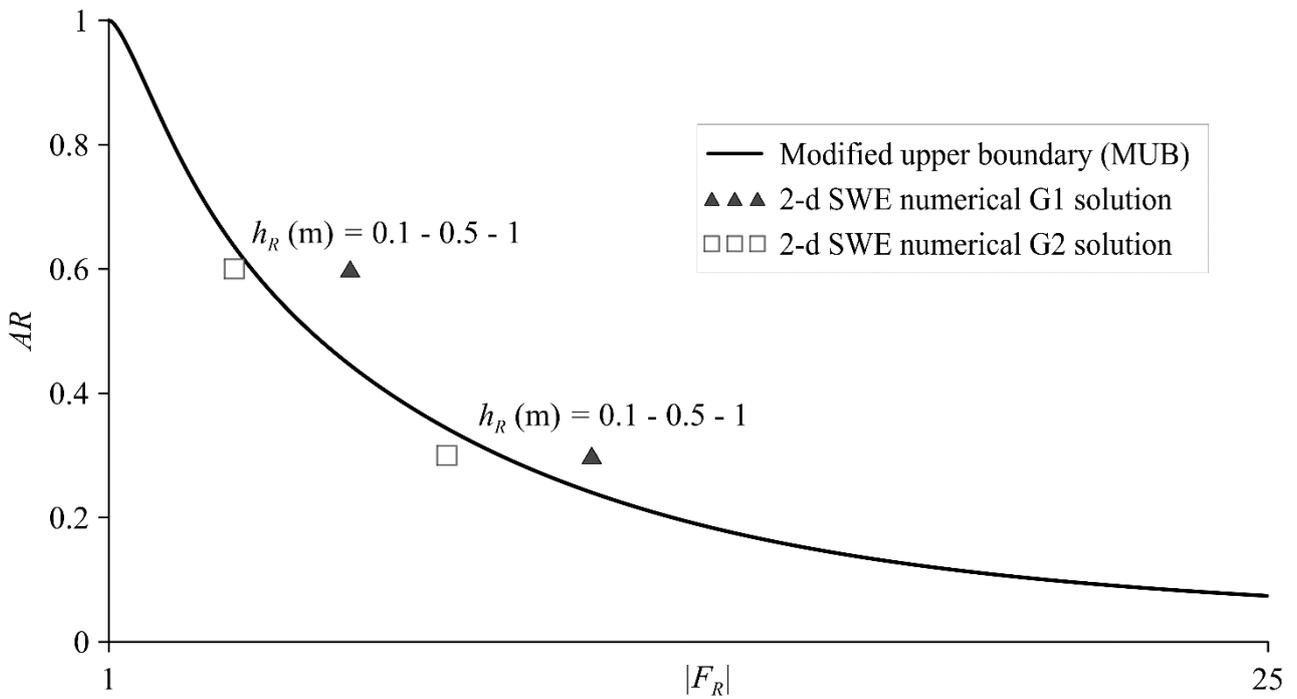

**Figure 22**. 2-d SWE numerical results for supercritical flows with $B_R = 1$ m and $h_R = 0.1$, 0.5, and 1 m impinging a contraction: G1 configuration (black triangles), G2 configuration (white squares); modified upper boundary (thick black line).

## 7. Conclusions

The solution of the Riemann problem associated to the Single Porosity (SP) Shallow water Equations (SWE) model by Guinot and Soares-Frazão (2006) is the main ingredient for the computation of interface fluxes and obstacle reaction terms in the Binary SP model (Varra et al. 2020) and in the integral approach (Sanders et al. 2008, Guinot et al. 2017). Previous studies (Cozzolino et al. 2018a, Varra et al. 2020, 2021) have shown that the SP Riemann problem presents a fundamental ambiguity consisting in the appearance of multiple exact solutions for certain initial conditions characterized by a supercritical flow impacting on a porosity reduction. This observation prompts the definition of the unique physically congruent Riemann solution among the alternatives and the construction of a numerical scheme able to reproduce this relevant solution.

Having recognized that the 1-d SP model is nothing but a crude simplification of the 2-d SWE model in a variable width rectangular channel, the channel analogy (Guinot and Soares-Frazão 2006, Sanders et al. 2008) has been exploited in this paper to disambiguate the multiple 1-d SP Riemann solutions by means of a systematic comparison with the corresponding 2-d SWE numerical solutions at local geometric discontinuities. The conclusion, which corresponds to other similar results obtained in the literature for different mathematical models (Han et al. 2013, Cozzolino et al. 2014, Aleksyuk and Belikov 2019), is that the solution with a backwards moving shock is physically congruent when multiple solutions are possible. Laboratory (Akers and Bokhove 2008, Defina and Viero 2010) and 2-d SWE numerical experiments (Varra et al. 2020) show that supercritical flows in channels suffer intense head loss through a width contraction, which corresponds to a porosity reduction. This phenomenon is an additional cause of energy dissipation in porosity models with respect to those already described in the literature (Guinot et al. 2017, 2018). Also in this case, the systematic study of 2-d SWE numerical results at isolated geometric discontinuities has supplied the general conditions under which this energy dissipation is present and how it can be evaluated.

Based on a modification of the generalized hydrostatic reconstruction by Castro et al. (2007), we have also built an approximate Riemann solver that discriminates the existence of multiple solutions and is able to add adequate head loss in the case of supercritical flow through porosity discontinuities. The comparison between the numerical results supplied by the novel 1-d SP model and the 2-d SWE model shows that the former can reproduce the effects that in 2-d models are caused by the interaction between a supercritical flow and a contraction. This promising numerical approach could be extended to other cases of hyperbolic systems of differential equations where multiple solutions arise such as the SWE and the porous SWE with variable topography.

**Aknowledgements**

Renata Della Morte and Luca Cozzolino want to acknowledge the financial support from the project "Floods in cities: new insights for integrating pluvial flooding into flood risk management plans (INSPIRING)", funded by the Italian Ministry of University and Research under the national programme PRIN2020.

**Appendix A. Head-balance form of the porosity discontinuity**

If $Q = \varphi_L h_1 u_1 = \varphi_R h_2 u_2$ is the unit-width discharge flowing through the porosity discontinuity, the velocities at the two sides of the geometric transition can be rewritten as

(A.1) $\quad u_1 = \dfrac{Q}{\varphi_L h_1}, \quad u_2 = \dfrac{Q}{\varphi_R h_2}$ .

The substitution of Eq. (A.1) into Eq. (5.b) leads to

(A.2) $\dfrac{g}{2}\left[\varphi_R h_2^2 - \varphi_L h_1^2\right] + \dfrac{Q^2}{\varphi_R h_2} - \dfrac{Q^2}{\varphi_L h_1} = S_\Gamma\left(\varphi_L, \varphi_R, \mathbf{u}_1, \mathbf{u}_2\right),$

while the substitution into the second of Eq. (6) supplies

(A.3) $\Delta H\left(\varphi_L, \varphi_R, \mathbf{u}_1, \mathbf{u}_2\right) = h_2 + \dfrac{Q^2}{2g\left(\varphi_R h_2\right)^2} - \left[h_1 + \dfrac{Q^2}{2g\left(\varphi_L h_1\right)^2}\right].$

The elimination of $Q^2$ between Eqs. (A.2) and (A.3) finally supplies Eq. (7).

**Appendix B. Smooth stationary weak solutions of Eq. (2)**

If the time derivatives are null, Eq. (2) can be rewritten as

(B.1) $\begin{aligned} &\dfrac{d\varphi h u}{ds} = 0 \\ &\dfrac{d}{ds}\left(\varphi \dfrac{gh^2}{2}\right) + \dfrac{d\varphi h u^2}{ds} - \dfrac{gh^2}{2}\dfrac{d\varphi}{ds} = 0 \end{aligned}.$

If the porosity and the flow variables are smooth (i.e., continuous with their derivatives), the second of Eq. (B.1) can be rewritten as

(B.2) $\varphi g h \dfrac{dh}{ds} + \dfrac{gh^2}{2}\dfrac{d\varphi}{ds} + \varphi h \dfrac{d}{ds}\left(\dfrac{u^2}{2}\right) + u\dfrac{d\varphi h u}{ds} - \dfrac{gh^2}{2}\dfrac{d\varphi}{ds} = 0.$

Finally, the substitution of the first of Eq. (B.1) into Eq. (B.2) and the cancellation of the terms with opposite sign leads to

(B.3) $\dfrac{d}{ds}\left(h+\dfrac{u^2}{2g}\right)=0$,

which states that the total head is uniform.

**Appendix C. Discussion of Eq. (10) and of the corresponding Froude limits**

This Appendix reports in a condensed form the discussion present in classic (Yarnell 1934, Chow 1959) and recent (Defina and Susin 2006, Castro et al. 2007, Akers and Bokhove 2008, Varra et al. 2021) literature with reference to 1-d flows in channels with variable width. Given the formal analogy between 1-d SWE model with variable width and 1-d porous SWE model, the discussion can be extended to porous models where the porosity symbol replaces the width symbol while the discharge is intended as unit-width discharge.

The parametric family of the flow states $\mathbf{u}=(h \;\; hu)^T$ with given width $B$ and discharge $Q=Bhu$ is defined by $\mathbf{u}(Q,B,h)=(h \;\; Q/B)^T$, where the flow depth $h$ is the parameter. The total head corresponding to the states of this family is a function of the parameter $h$ only, and it is defined by

(C.1) $H(Q,B,h)=H(\mathbf{u}(Q,B,h))=h+\dfrac{Q^2}{2g(Bh)^2}$.

The function $H(Q, B, h)$ in Eq. (C.1) is convex with respect to $h > 0$, it is positive and it has a unique minimum in $h=h_c(Q,B)$. The critical depth $h_c(Q,B)$ and the corresponding critical head $H_c(Q,B)=H(Q,B,h_c(Q,B))$ are defined by (Chow 1959)

(C.2) $h_c(Q,B) = \sqrt[3]{\dfrac{Q^2}{gB^2}}, \quad H_c(Q,B) = \dfrac{3}{2}\sqrt[3]{\dfrac{Q^2}{gB^2}}.$

The states **u** with $h < h_c(Q,B)$ are characterized by $|F(\mathbf{u})| > 1$ and are called supercritical. The states with $h > h_c(Q,B)$ are characterized by $0 < |F(\mathbf{u})| < 1$ and are called subcritical. From the preceding discussion, it follows that a state **u** with discharge $Q$ has energy sufficient to pass through the cross-section whose width is $B$ only if the corresponding head is not minor than $H_c(Q,B)$.

This observation has consequences for the application of Eq. (10), expressing the invariance of discharge and head. Let $B_L$ and $B_R$ be the channel widths at the left and right ends, respectively, of a geometric transition, and let $\mathbf{u}_2 = (h_2 \quad h_2 u_2)^T$ be the state corresponding to the flow at the right end. It is possible to find the left end state $\mathbf{u}_1 = (h_1 \quad h_1 u_1)^T$ connected to $\mathbf{u}_2$ by means of the discharge and head invariance only if $H(\mathbf{u}_1) \geq H_c(Q, B_L)$, namely only if

(C.3) $H(\mathbf{u}_1) \geq \dfrac{3}{2}\sqrt[3]{\dfrac{Q^2}{gB_L^2}},$

where $Q = B_L h_1 u_1$ is the discharge corresponding to the right state $\mathbf{u}_1$. The invariance of discharge and head is expressed by $B_R h_2 u_2 = B_L h_1 u_1$ and $H(\mathbf{u}_2) = H(\mathbf{u}_1)$, implying that the condition of Eq. (C.3) can be rewritten as

(C.4) $H(\mathbf{u}_2) \geq \dfrac{3}{2}\sqrt[3]{\dfrac{1}{AR^2}\dfrac{(h_2 u_2)^2}{g}},$

where $AR = B_L/B_R$ is the aspect ratio. If the Eq. (C.4) is satisfied, it exists the state $\mathbf{u}_1$ connected to $\mathbf{u}_2$ by the invariance of discharge and head with aspect ratio $AR$.

The Eq. (C.4) can be rewritten in dimensionless form as (Yarnell 1934, Chow 1959, Defina and Susin 2006, Akers and Bokhove 2008)

(C.5) $AR \geq f\left(\left|F(\mathbf{u}_2)\right|\right),$

where $F(\mathbf{u}_2) = u_2/\sqrt{gh_2}$ is the Froude number of the state $\mathbf{u}_2$ and the function $f(x)$ is defined as

(C.6) $f(x) = x\left(\dfrac{3}{2+x^2}\right)^{3/2}, \quad x \geq 0.$

The function $f(x)$ of Eq. (C.6) is non-negative, strictly increasing for $x \in [0,1[$, strictly decreasing for $x > 1$, and it has a maximum in $x = 1$ with $f(1) = 1$. The properties of the function $f(x)$ have the following implications.

*Case $AR > 1$.* In case of $AR > 1$, Eq. (C.5) is satisfied for every $F(\mathbf{u}_2)$. In other words, it always exists the state $\mathbf{u}_1$ connected to $\mathbf{u}_2$ by the invariance of discharge and head when there is a width increase.

*Case $AR \leq 1$.* In case of $AR \leq 1$, Eq. (C.5) is satisfied by

(C.7) $\left|F(\mathbf{u}_2)\right| \leq K_{sb}(AR), \quad \left|F(\mathbf{u}_2)\right| \geq K_{sp}(AR),$

where the limit Froude numbers $K_{sb}(AR)$ and $K_{sp}(AR)$ are defined as (Varra et al. 2021)

$$K_{sp}(AR) = (AR)^{-1/2} \left[ 2\cos\left(\frac{\pi}{3} - \frac{1}{3}\arctan\sqrt{(AR)^{-2}-1}\right) \right]^{\frac{3}{2}}$$

(C.8)

$$K_{sb}(AR) = (AR)^{-1/2} \left[ 2\cos\left(\frac{5\pi}{3} - \frac{1}{3}\arctan\sqrt{(AR)^{-2}-1}\right) \right]^{\frac{3}{2}}.$$

The Froude limits defined by Eq. (C.8) are characterised by $K_{sp}(AR) \geq 1$ (supercritical) and $K_{sb}(AR) \leq 1$ (subcritical) for every $AR \leq 1$.

In conclusion, the present discussion shows that two different situations are possible with reference to Eq. (10) where $AR = \varphi_L/\varphi_R \leq 1$:

a) Given the state $\mathbf{u}_1$, it is always possible to find the corresponding state $\mathbf{u}_2$.

b) Given the state $\mathbf{u}_2$, it is possible to find the corresponding state $\mathbf{u}_1$ only if one of the two inequalities of Eq. (C.7) is satisfied. In this case, the invariance of the total head through the porosity transition implies that $\mathbf{u}_1$ is subcritical [supercritical] if $\mathbf{u}_2$ is subcritical [supercritical], and *vice versa*. When $|F(\mathbf{u}_2)| = K_{sb}(AR)$ or $|F(\mathbf{u}_2)| = K_{sp}(AR)$, the state $\mathbf{u}_1$ is critical.

For $AR = \varphi_L/\varphi_R \leq 1$, it is possible to define an additional limit $K_{jump}(AR)$ for the Froude number, as follows. Let the left state $\mathbf{u}_1$ be critical and the right state $\mathbf{u}_2$ be subcritical with $F(\mathbf{u}_2) = -K_{sb}(AR)$ (flow from right to left). The supercritical state $\mathbf{u}_2^\#$ to the right of the subcritical state $\mathbf{u}$ and connected to it by a standing hydraulic jump is characterized by Froude number $F(\mathbf{u}_2^\#) = -K_{jump}(AR)$, where

(C.9) $K_{jump}(AR) = K_{sb}(AR)\sqrt{8}\left(-1 + \sqrt{1 + 8K_{sb}^2(AR)}\right)^{-\frac{3}{2}}.$

## Appendix D. Head loss through a standing hydraulic jump

The flow depth $h_R^\#$ corresponding to the state $\mathbf{u}_R^\# = \begin{pmatrix} h_R^\# & h_R^\# u_R^\# \end{pmatrix}^T$ connected to $\mathbf{u}_R$ by means of a hydraulic jump in a rectangular channel is (Chow 1959)

(D.1) $h_R^\# = \dfrac{h_R}{2}\left(-1 + \sqrt{1 + 8F_R^2}\right),$

where $F_R^2$ is the squared Froude number corresponding to the state $\mathbf{u}_R$. From Eq. (D.1), the ratio $h_R^\#/h_R$ depends on $F_R^2$ only.

The discharge is conserved through the hydraulic jump, implying that $h_R^\# u_R^\# = h_R u_R$. For this reason, the head $H_R^\#$ corresponding to the state $\mathbf{u}_R^\#$ is

(D.2) $H_R^\# = H\left(\mathbf{u}_R^\#\right) = h_R^\# + \dfrac{u_R^2}{2g}\left(\dfrac{h_R}{h_R^\#}\right)^2 = h_R^\#\left[1 + \dfrac{F_R^2}{2}\left(\dfrac{h_R}{h_R^\#}\right)^3\right].$

Once that Eq. (D.1) is substituted into Eq. (D.2), the relative head loss $\Delta^\# = \Delta H^\#/H_R = \left(H_R - H_R^\#\right)/H_R$ can be easily calculated as

(D.3) $\Delta^\# = \dfrac{H_R - H_R^\#}{H_R} = 1 - \dfrac{h_R^\#}{h_R}\left[1 + \dfrac{F_R^2}{2}\left(\dfrac{h_R}{h_R^\#}\right)^3\right]\left[1 + \dfrac{F_R^2}{2}\right]^{-1},$

where $H_R = H\left(\mathbf{u}_R\right)$ is the head corresponding to the state $\mathbf{u}_R$. From Eqs. (D.1) and (D.3), it is evident that the relative head loss $\Delta^\#$ through the hydraulic jump depends on $F_R^2$ only.

In the case that the supercritical state $\mathbf{u}_R$ is connected to the subcritical state $\mathbf{u}_2$ by a hydraulic jump (i.e., when $\mathbf{u}_2 = \mathbf{u}_R^\#$) and $\mathbf{u}_1$ is critical (see Figure 4c), $F_R^2$ coincides with $K_{jump}^2(AR)$ (see Appendix C) and the relative head loss $\Delta^\#$ depends on $AR$ only.

**References**

Akers B., Bokhove O. (2008) Hydraulic flow through a channel contraction: multiple steady states, Physics of Fluids 20, 056601. Doi: 10.1063/1.2909659.

Aleksyuk A.I., Belikov V.V. (2019) The uniqueness of the exact solution of the Riemann problem for the shallow water equations with discontinuous bottom, Journal of Computational Physics 390, 232-248. Doi: 10.1016/j.jcp.2019.04.001.

Aleksyuk A.I., Malakhov M.A., Belikov V.V. (2022) The exact Riemann solver for the shallow water equations with a discontinuous bottom, Journal of Computational Physics 450, 110801. Doi: 10.1016/j.jcp.2021.110801.

Austin L.H., Skogerboe G.V., Bennett R.S. (1970) Subcritical flow at open channel structures. Open channel expansions. Utah State University Reports, 638. https://digitalcommons.usu.edu/water_rep/638.

Bruwier M., Archambeau P., Erpicum S., Pirotton M., Dewals B. (2017) Shallow-water models with anisotropic porosity and merging for flood modelling on Cartesian grids, Journal of Hydrology 554, 693-709. Doi: 10.1016/j.jhydrol.2017.09.051.

Buckingham E. (1914) On physically similar systems; illustrations of the use of dimensional equations. Physical Review 4(4), 345-376.

Castro M.J., Pardo Milanés A., Parés C. (2007) Well-balanced numerical schemes based on a generalized hydrostatic reconstruction technique, Mathematical Models and Methods in Applied Sciences 17(12), 2055-2113. Doi: 10.1142/S021820250700256X.

Cea L., Vázquez-Cendón M.E. (2010) Unstructured finite volume discretization of two-dimensional depth-averaged shallow water equations with porosity, International Journal for Numerical Methods in Fluids 63(8), 903-930. http://dx.doi.org/10.1002/fld.2107.

Chow V.T. (1959) Open channel hydraulics. McGraw-Hill, New York.

Cozzolino L., Castaldo R., Cimorelli L., Della Morte R., Pepe V., Varra G., Covelli C., Pianese D. (2018a) Multiple solutions for the Riemann problem in the Porous Shallow water Equations, EPiC Series in Engineering 3, 476-484. Doi: 10.29007/31n4.

Cozzolino L., Cimorelli L., Covelli C., Della Morte R., Pianese D. (2014) Boundary conditions in finite volume schemes for the solution of shallow-water equations: the non-submerged broad-crested weir, Journal of Hydroinformatics 14(6), 1235-1249. Doi: 10.2166/hydro.2014.100.

Cozzolino L., Pepe V., Cimorelli L., D'Aniello A., Della Morte R., Pianese D. (2018b) The solution of the dam-break problem in the Porous Shallow water Equations, Advances in Water Resources 114, 83-101. Doi: 10.1016/j.advwatres.2018.01.026.

Cozzolino L., Pepe V., Morlando F., Cimorelli L., D'Aniello A., Della Morte R., Pianese D. (2017) Exact solution of the dam-break problem for constrictions and obstructions in constant width rectangular channels, ASCE Journal of Hydraulic Engineering 143(11), 04017047. https://doi.org/10.1061/(ASCE)HY.1943-7900.0001368.

Cunge J.A., Holly H.M., Verwey A. (1980) Practical aspects of computational river hydraulics, Pitman, Boston.

Dal Maso G., LeFloch P.G., Murat F. (1995) Definition and weak stability of nonconservative products, Journal del Mathématiques Pures et Appliqués 74(6), 483-548.

Defina A. (2000) Two-dimensional shallow flow equations for partially dry areas, Water Resources Research 36(11), 3251-3264. http://dx.doi.org/10.1029/2000WR900167.

Defina A., Susin F.M. (2006) Multiple states in open channel flow, in *Vorticity and turbulence effects in fluids structures interactions – Advances in Fluid Mechanics*, M. Brocchini and F. Trivellato eds., Wessex Institute of Technology Press, Southampton, 105-130.

Defina A., Viero D.P. (2010) Open channel flow through a linear contraction, Physics of Fluids 22(3), 036602. Doi: 10.1063/1.3370334.

Dewals B., Bruwier M., Pirotton M., Erpicum S., Archambeau P. (2021) Porosity Models for Large-Scale Urban Flood Modelling: A Review, Water, 13(7), 960. https://doi.org/10.3390/w13070960.

Ferrari A., Vacondio R., Dazzo S., Mognosa P. (2017) A 1d–2d shallow water equations solver for discontinuous porosity field based on a generalized Riemann problem, Advances in Eater Resources 107, 233-249. Doi: 10.1016/j.advwatres.2017.06.023

Finaud-Guyot P., Delenne C., Lhomme J., Guinot V., Llovel C. (2010) An approximate-state Riemann solver for the two-dimensional shallow water equations with porosity, International Journal for Numerical Methods in Fluids 62(12), 1299-1331. Doi: 10.1002/fld.2066.

Formica G. (1955) Esperienze preliminari sulle perdite di carico nei canali dovute a cambiamenti di sezione, L'Energia Elettrica 32(7), 554-567 (in Italian).

Godlewski E., Raviart P.-A. (1996) Numerical approximation of systems of hyperbolic equations, Springer, New York.

Guinot V., Delenne C. (2014) Macroscopic modelling of urban floods, La Houille Blanche 6, 19-25. Doi: 10.1051/lhb/2014058.

Guinot V., Delenne C., Rousseau A., Boutron O. (2018) Flux closures and source terms models for shallow water models with depth-dependent integral porosity, Advances in Water Resources 122, 1-26. Doi: 10.1016/j.advwatres.2018.09.014.

Guinot V., Delenne C., Soares-Frazão S. (2022) Self-similar solutions of shallow water equations with porosity, IAHR Journal of Hydraulic Research. Doi: 10.1080/00221686.2022.2106598.

Guinot V., Sanders B.F., Schubert J.E. (2017) Dual integral porosity shallow water model for urban flood modelling, Advances in Water Resources 103, 16-31. Doi: 10.1016/j.advwatres.2017.02.009.

Guinot V., Soares-Frazão S. (2006) Flux and source term discretization in two-dimensional shallow water models with porosity on unstructured grids, International Journal for Numerical Methods in Fluids 50(3), 309-345. Doi: 10.1002/fld.1059.

Hager W. (2010) Wastewater hydraulics: theory and practice. Springer, Berlin.

Hager W. H., Sinniger R. (1985) Flow characteristics of the hydraulic jump in a stilling basin with an abrupt bottom rise. Journal of Hydraulic Research 23(2), 101–113. Doi: 10.1080/00221688509499359.

Han E., Handke M., Warnecke G. (2013) Criteria for non-uniqueness of Riemann solutions to compressible duct flows, Zeitschrift für Angewandte Mathematik und Mechanik 93(6-7), 465-475. Doi: 10.1002/zamm.201100176.

Han E., Warnecke G. (2014) Exact Riemann solutions to Shallow water Equations, Quarterly of Applied Mathematics 72(3), 407-453. Doi: 10.1090/S0033-569X-2014-01353-3.

Ion S., Marinescu D., Ion A.V., Cruceanu S.G. (2022) Numerical scheme for solving a porous Saint-Venant type model for water flow on vegetated hillslopes, Applied Numerical Mathematics 172, 67-98. Doi: 10.1016/j.apnum.2021.09.019.

Ippen A.T., Dawson J.H. (1951) High-velocity flow in open channels: a symposium: Design of Channel Contractions. ASCE Transactions 116(1), 326–346. https://doi.org/10.1061/TACEAT.0006522.

Jung J. (2022) Development of exact solution and finite-volume method of porous shallow water equations for the analysis of urban flooding, Seoul National University, PhD dissertation. https://hdl.handle.net/10371/181177.

Karki K. S., Chander S., Malhotra, R. C. (1972) Supercritical flow over sills at incipient jump conditions, ASCE Journal of the Hydraulic Division 98(10), 1753–1764. Doi: 10.1061/JYCEAJ.0003435.

LeFloch P. (1989) Shock waves for nonlinear hyperbolic systems in nonconservative form, Institute for Mathematics and Its Applications Preprint Series 593, Minneapolis. http://hdl.handle.net/11299/5107.

Lhomme J. (2006) Modelisation des inondations en milieu urbain: approaches unidimensionelle, bidimensionnelle et macroscopique. Hydrologie. PhD Thesis, Universite Montpellier II – Sciences e Techniques du Languedoc. (in French).

Mohamed K. (2014) A finite volume method for numerical simulation of shallow water models with porosity, Computer & Fluids 104, 9-19. Doi: 10.1016/j.compfluid.2014.07.020.

Özgen I., Liang D.-f., Hinkelmann R. (2016a) Shallow water equations with depth-dependent anisotropic porosity for subgrid-scale topography, Applied Mathematical Modelling 40(17-18), 7447-7473. Doi: 10.1016/j.apm.2015.12.012.

Özgen I., Zhao J.-h., Liang D.-f., Hinkelmann R. (2016b) Urban flood modeling using shallow water equations with depth-dependent anisotropic porosity, Journal of Hydrology 541(Part B), 1165-1184. Doi: 10.1016/j.jhydrol.2016.08.025.

Özgen I., Zhao J.-h., Liang D.-f., Hinkelmann R. (2017) Wave propagation speeds and source term influences in single and integral porosity shallow water equations, Water Science and Engineering 10(4), 275-286. Doi: 10.1016/j.wse.2017.12.003.

Sanders B.F., Schubert J.E., Gallegos H.A. (2008) Integral formulation of shallow-water equations with anisotropic porosity for urban flood modelling, Journal of Hydrology 362(1-2), 19-38. Doi: 10.1016/j.jhydrol.2008.08.009.

Soares-Frazão S., Lhomme J., Guinot V., Zech Y. (2008) Two-dimensional shallow water model with porosity for urban flood modelling, Journal of Hydraulic Research 46(1), 45-64.

Valiani A., Caleffi V. (2008) Depth-energy and depth-force relationships in open channel flows: Analytical findings. Advances in Water Resources 31(3), 447-454. Doi: 10.1016/j.advwatres.2007.09.007.

Varra G., Pepe V., Cimorelli L., Della Morte R., Cozzolino L. (2020) On integral and differential porosity models for urban flooding simulation. Advances in Water Resources 136. Doi: 10.1016/j.advwatres.2019.103455.

Varra G., Pepe V., Cimorelli L., Della Morte R., Cozzolino L. (2021) The exact solution to the Shallow water Equations Riemann problem at width jumps in rectangular channels. Advances in Water Resources 155. Doi: 10.1016/j.advwatres.2021.103993.

Varra G., Della Morte R., Gargano R., Cozzolino L. (2022) Porous Shallow water Equations model with disambiguation of multiple solutions, Environmental Sciences Proceedings 21(1), 55. Doi: 10.3390/environsciproc2022021055.

Vaschy A. (1892) Sur les lois de similitude en physique, Annales Tèlègraphiques 19, 25-28.

Viero D.P., Defina A. (2017) Extended theory of hydraulic hysteresis in open-channel flow, Journal of Hydraulic Engineering 143(9), 06017014. Doi: 10.1061/(ASCE)HY.1943-7900.0001342.

Whitaker S. (1969) Advances in the theory of fluid motion in porous media, Industrial and Engineering Chemistry 61(12), 14-28. Doi: 10.1021/ie50720a004.

Yarnell D.L. (1934) Bridge piers as channel obstructions, Technical Bulletin 444, US Department of Agriculture, Washington. https://naldc.nal.usda.gov/download/CAT86200436/PDF.

**Tables List**

Table 1. Initial flow conditions of the validation Riemann problems.

Table 2. Coefficients for the polynomial interpolation of Eq. (11).

Table 3. Coefficients for the polynomial interpolation of Eq. (12).

Table 4. Supercritical 2-d flow impinging on a contraction for $B_R = 1$ m and $h_R = 0.1, 0.5$, and 1 m: relative head losses $\Delta^*$ for the cases of flow passing through the discontinuity (G1). A hyphen indicates the cases where a backwards moving shock is produced (G2 configurations).

**Figures List**

Figure 1. Physical interpretation of the porosity discontinuity between $\varphi_L$ and $\varphi_R$: monotonic (a) and non-monotonic porosity variation (b).

Figure 2. Internal description of the porosity discontinuity: plan view of the monotonic porosity variation (a); profile view of smooth flow depth variation (b); profile view of flow depth variation with hydraulic jump (c).

Figure 3. Field of occurrence of multiple solutions to the porosity Riemann problem for right supercritical flows $\mathbf{u}_R$ impinging a porosity reduction with $AR = \varphi_L/\varphi_R \leq 1$. Lower (continuous line) and upper (dashed line) boundaries of the hysteresis domains. Hysteresis domains: A (no multiple solutions), B (multiple solutions even in the case $h_L = 0$), C (multiple solutions only for $h_L > 0$).

Figure 4. Flow conditions through the porosity discontinuity when multiple solutions to the purely 1-d SP Riemann problem are possible: profile view of solutions T1 (a), T2 (b) and T3 (c).

Figure 5. Plan view of the channel considered for 2-d SWE numerical simulations. Distorted representation (measures in metres).

Figure 6. Profile view of the 1-d SP exact (continuous black line) and 2-d SWE numerical solutions (dots) for the flow depth at time $t = 5$ s. Example Riemann problems 1 (a), 2 (b), 3 (c) and 4 (d) with initial conditions in Table 1.

Figure 7. Plan view of the 2-d SWE numerical solution for Riemann problem 3 with initial conditions in Table 1. Flow depth contours at time $t = 5$ s.

Figure 8. Example Riemann problem 5 with initial conditions in Table 1. Profile view for the flow depth solution at time $t = 5$ s. 1-d SP exact solutions T1 (a), T2 (b) and T3 (c). Comparison between the T3 exact solution (continuous line) and the 2-d SWE numerical solution (dots) (d).

Figure 9. Example Riemann problem 6 with initial conditions in Table 1. Profile view for the flow depth solution at time $t = 5$ s. 1-d SP exact solutions T1 (a), T2 (b) and T3 (c). Comparison between the T3 exact solution (continuous line) and the 2-d SWE numerical solution (dots) (d).

Figure 10. Profile view of the 1-d SP exact (continuous black line) and 2-d SWE numerical solutions (dots) for the flow depth at time $t = 5$ s. Example Riemann problems 7 (a) and 8 (b) with initial conditions in Table 1.

Figure 11. Plan view of the 2-d SWE numerical solution for Riemann problem 7 with initial conditions in Table 1. Flow depth contours at time $t = 5$ s.

Figure 12. Plan view of the 2-d SWE numerical solution for Riemann problem 8 with initial conditions in Table 1. Flow depth contours at time $t = 5$ s.

Figure 13. Plane view of the channel used for 2-d SWE numerical tests with supercritical flows. Distorted representation (measures in metres).

Figure 14. 2-d SWE numerical results for supercritical flows with $B_R = 1$ m and $h_R = 1$ m impinging a contraction: G1 configuration (black triangles), G2 configuration (white squares); upper hysteresis domain limit (dashed line); modified upper boundary (thick black line).

Figure 15. Relative head losses for supercritical flows through a contraction: 2-d SWE numerical results for G1 configuration (black triangles); limit relative head loss (continuous black line); envelope of G1 data closer to the modified upper boundary of Figure 14 (continuous grey line).

Figure 16. Polynomial interpolation of the relative head loss data. Triangles represent the experimental cases enveloped by a thin grey line in Figure 15.

Figure 17. Profile view of the 1-d SP exact solution with head loss through the geometric discontinuity (continuous black line) and 2-d SWE numerical solution (dots) for the flow depth at time $t = 5$ s. Example Riemann problems 7 (a) and 8 (b) with initial conditions in Table 1.

Figure 18. Side view of two neighbouring cells in the 1-d computational domain: cell-averaged quantities at a generic time level $n$ (a); interface reconstructed variables used in the basic reconstruction by Castro et al. (2007) (b); interface and in-cell reconstructed variables in the novel reconstruction approach (c).

Figure 19. Profile view of the numerical solution for the flow depth at time $t = 5$ s: 1-d SP model with the variable reconstruction by Castro et al. (2007) (continuous black line) and 2-d SWE model (dots). Example Riemann problems 1 (a), 2 (b), 3 (c) and 4 (d) with initial conditions in Table 1.

Figure 20. Profile view of the numerical solution for the flow depth at time $t = 5$ s: 1-d SP model with the basic variable reconstruction by Castro et al. (2007) (continuous black line) and 2-d SWE model (dots). Example Riemann problems 5 (a), 6 (b), 7 (c) and 8 (d) with initial conditions in Table 1.

Figure 21. Profile view of the numerical solution for the flow depth at time $t = 5$ s: 1-d SP model with the novel variable reconstruction (continuous black line) and 2-d SWE model (dots). Example Riemann problems 5 (a), 6 (b), 7 (c) and 8 (d) with initial conditions in Table 1.

Figure 22. 2-d SWE numerical results for supercritical flows with $B_R = 1$ m and $h_R = 0.1, 0.5$, and 1 m impinging a contraction: G1 configuration (black triangles), G2 configuration (white squares); modified upper boundary (thick black line).